%% file: main.tex
\documentclass[pdflatex,sn-basic]{sn-jnl}

\usepackage[T1]{fontenc}%
\usepackage{newtxtext,newtxmath}%

\usepackage{pdflscape}
\usepackage{ragged2e}
\usepackage{graphicx}%
\usepackage{longtable}
\usepackage{array}
\usepackage{multirow}%
\usepackage{booktabs}%
\usepackage{tabularx}%
\usepackage{amsmath,amsfonts}%
\usepackage{enumitem}

\graphicspath{{figures/}{./}}

\newlist{cellenum}{enumerate}{1}

\setlist[cellenum]{
  label=\arabic*.,
  leftmargin=*,
  nosep,
  topsep=0pt,
  partopsep=0pt,
  parsep=0pt,
  before=\vspace{-0.35\baselineskip}
}

\AtBeginDocument{
\hypersetup{
    hidelinks,
    pdftitle={From Community Values to AI Design Constraints: A Mixed-Methods Study of a Proposed AI-Driven Wildfire Risk Assessment Tool in Los Angeles County},
    pdfauthor={Sanaz Sadat Hosseini, Mona Azarbayjani, Mohammad Pourhomayoun, Hamed Tabkhi}
}
}

\begin{document}

%
%

\vspace*{0.15in}

\begin{center}


{\fontsize{17.28}{20.5}\selectfont\bfseries
From Community Values to AI Design Constraints:
\\
A Mixed-Methods Study of a Proposed AI-Driven Wildfire Risk
\\
Assessment Tool in Los Angeles County
\par}

\vspace{1.90em}

{\large
Sanaz Sadat Hosseini\textsuperscript{1}
\quad
Mona Azarbayjani\textsuperscript{2}
\quad
Mohammad Pourhomayoun\textsuperscript{3}
\quad
Hamed Tabkhi\textsuperscript{4}
\par}

\vspace{2.0em}


{\small
\textsuperscript{1}Department of Civil and Environmental Engineering,
University of North Carolina at Charlotte, Charlotte, NC, USA
\par}

\vspace{0.10em}

{\small
\textsuperscript{2}David R. Ravin School of Architecture,
University of North Carolina at Charlotte, Charlotte, NC, USA
\par}

\vspace{0.10em}

{\small
\textsuperscript{3}Department of Computer Science,
California State University, Los Angeles, Los Angeles, CA, USA
\par}

\vspace{0.10em}

{\small
\textsuperscript{4}Department of Electrical and Computer Engineering,
University of North Carolina at Charlotte, Charlotte, NC, USA
\par}

\end{center}

\vspace{1.30em}

%

\begin{quotation}

\noindent
{\small\bfseries Abstract}

\vspace{0.35em}

\small
AI-driven hazard tools are increasingly being proposed for resident-facing risk communication. However, residents are often asked for feedback only after decisions about data use, privacy, and explanation have already been made. This study takes an earlier approach by examining residents' values, concerns, and practical expectations related to a proposed AI-based risk assessment tool and its anticipated outputs before a predictive model or functional interface is developed. The proposed smartphone application would use exterior property photographs to estimate parcel-level wildfire risk and provide an interpretable risk score, uncertainty information, and mitigation recommendations. Using a convergent mixed-methods design, we surveyed 30 Los Angeles County residents, seven of whom also participated in a structured virtual town hall. The county was selected as the study setting because of its high wildfire exposure and recurring fire risk. We analyzed the data using descriptive statistics, exploratory FDR-adjusted Spearman correlations, and inductive thematic coding of open-ended responses and town hall discussion. Findings indicate that participants were cautiously receptive to the proposed tool: 66.7\% said they would be very likely or likely to use the tool. They linked fair risk assessment to whether the tool considered relevant property and neighborhood conditions. Privacy and data security were the most common concerns (64.3\%), while cost was the main barrier to acting on recommendations (65.5\%). A strong association was found between participants’ likelihood of using the tool and their likelihood of saving or sharing their risk results {\normalfont\unboldmath ($\rho = 0.752$, $p_{\mathrm{FDR}} < .001$).} We translated the combined findings into the \emph{AI Value Map}, a community-grounded artifact organized around nine value dimensions. The Map connects participant findings to provisional design requirements and identifies possible consequences if those requirements are not addressed. This study contributes exploratory evidence on residents’ responses to a proposed AI wildfire tool and offers a pre-model procedure for translating early community input into traceable design constraints for resident-facing AI hazard tools.

\end{quotation}


\vspace{-0.30em}

\begin{quotation}
\small
\noindent
\textbf{Keywords:}
Participatory AI;
Explainable AI;
Wildfire Risk Communication;
Human-Centered AI;
Value-Sensitive Design;
Community Engagement
\end{quotation}

\vspace{0.80em}

%

\input{sections/01_introduction}

\input{sections/02_background}

\input{sections/03_study_framing}

\input{sections/04_methods}

\input{sections/05_findings}

\input{sections/06_ai_value_map}

\input{sections/07_discussion}

\input{sections/08_limitations_future_work}

\input{sections/09_conclusion}


\backmatter

\setlength{\bibsep}{0pt}
\bibliography{references}

\end{document}

%% file: sections/01_introduction.tex
\section{Introduction and Problem Context}
\label{sec:introduction}

Wildfires are a growing threat to communities worldwide, particularly in fire-prone regions where climate change, urban expansion, and inadequate mitigation together intensify wildfire risk \citep{IPCC2023,Radeloff2023,McCaffrey2015}. In the United States, particularly in California, wildfire has become a recurring feature of urban and peri-urban life. Longer droughts, hotter summers, and continued development in fire-prone landscapes have increased the likelihood that fires will ignite and spread rapidly \citep{AmericanPlanningAssociation2016}. Southern California illustrates this challenge clearly. The January 2025 Palisades and Eaton Fires destroyed more than sixteen thousand structures across affected communities, including Pacific Palisades, Altadena, and Pasadena, and showed how wildfire damage can vary sharply across nearby neighborhoods \citep{CALFIREEaton2025, CALFIREPalisades2025}.

\begin{figure}[ht]
\centering
\includegraphics[width=\textwidth]{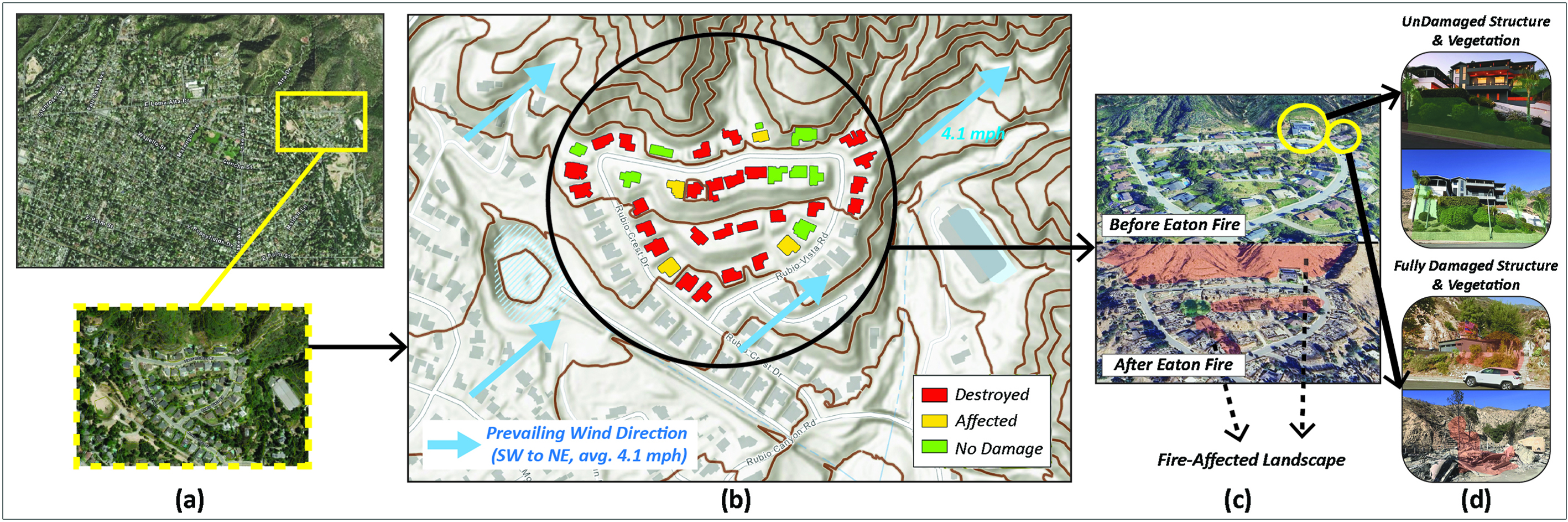}
\caption{Parcel-level variation in damage during the January 2025 Eaton Fire in Altadena, CA. (a) Regional context of the Altadena foothills area with a zoomed-in neighborhood area affected by the Eaton Fire; (b) Site-level map showing parcel-level fire damage classifications over topographic contours, with prevailing wind direction to contextualize fire behavior; (c) Pre- and post-fire aerial imagery highlighting uneven damage along a single street segment; (d) Ground-level photographs of two nearby homes in the foothill area showing contrasting outcomes and surrounding exterior conditions. Source: Authors’ own work.}\label{fig1}
\end{figure}

An important dimension of wildfire vulnerability is the spread of fire from one property to another through interconnected conditions in landscaping, exterior materials, and spacing between structures \citep{GonzalezMathiesenMarch2018}. Recent fire events have shown that one home may be destroyed while a neighboring property remains intact, with differences often related to defensible space, vegetation conditions, building materials, and parcel layout \citep{Penney2025}. For this reason, wildfire resilience cannot be understood only as an individual household responsibility. It also depends on shared, neighborhood-level awareness of risk and on coordinated mitigation across properties \citep{Stidham2014}. Figure \ref{fig1} illustrates parcel-level variation in damage and destruction during the January 2025 Eaton Fire.

Despite the importance of property-level conditions, residents may remain uncertain about what makes their homes more or less vulnerable \citep{Larsen2025}. Current public-facing tools provide important mapping and modeling capabilities, but major national resources such as \emph{Wildfire Risk to Communities} rely on nationally consistent modeled datasets and are not intended for locally calibrated, neighborhood-scale or individual-home-scale assessment \citep{USDAForestService2024}. Other technical products focus on burned-area mapping and provide limited resident-facing prevention guidance \citep{Giglio2018}. Residents may therefore lack the clear, property-specific information needed to understand risk and take appropriate action.

At the same time, AI-based wildfire risk assessment has advanced rapidly. Although these advances expand the analytic capacity available for mapping and estimating risk, they do not by themselves address the resident-facing communication gap \citep{Jain2020,Huot2022}. Existing explainable AI (XAI) tools \citep{Arya2019,Wexler2019}, ethical AI frameworks \citep{Floridi2018,Shneiderman2020}, and participatory design approaches \citep{Birhane2022,Sloane2022} provide important resources for explanation, accountability, and community participation. However, they offer limited guidance on how residents' values and concerns related to AI-generated assessments and outputs can be translated into specific design constraints before key technical and interface decisions are fixed. These include concerns about what data the system uses, how risk scores and uncertainty are generated and explained, how privacy and fairness are handled, how residents provide information and use the results, and whether recommendations are feasible in their circumstances. For resident-facing wildfire tools, addressing these issues early can shape whether residents understand the results, trust the system, and view its recommendations as feasible and relevant \citep{McCaffrey2015,Shneiderman2020}. These choices may also influence whether the tool supports equitable risk communication or reinforces existing disparities in wildfire vulnerability, preparedness, and recovery \citep{Davies2018,Ong2025}.

This study addresses this gap through an exploratory mixed-methods investigation of Los Angeles County residents. The study was motivated by the January 2025 Palisades and Eaton Fires and their impacts in Altadena, Pasadena, and Pacific Palisades cities in Los Angeles County. The tool examined in this study is a smartphone application that would use exterior property photographs submitted by residents as one input and integrate them with satellite imagery and remote-sensing observations, meteorological data, ground-based information, vegetation maps, and historical wildfire data to support a more accurate and context-sensitive parcel-level risk assessment and provide an interpretable risk score, uncertainty information, and property-specific mitigation recommendations. At the time of the study, the tool existed only as wireframes and conceptual flows; no property-specific predictive model had been developed. Using a survey ($n = 30$) and a structured virtual town hall meeting ($n = 7$), we examined residents' values, concerns, and practical expectations related to the proposed AI assessment, including its data inputs, anticipated risk outputs, uncertainty information, recommendations, privacy, and fairness. We then translated these findings into the \emph{AI Value Map}, which is an artifact organized around nine value dimensions. Although these dimensions were derived for the proposed wildfire tool examined here, the evidence-to-constraint procedure used to build the Value Map may also be applied in other AI hazard settings, where different community findings, dimensions, and constraints would be expected to emerge. Accordingly, this study addresses three research questions:
        
\begin{itemize}
\item \emph{RQ1.} What values, concerns, and practical barriers do residents associate with a proposed photo-based AI wildfire risk assessment tool?
\item \emph{RQ2.} How do participants interpret key features of the conceptual wireframes and materials, including risk scores, explanations, privacy, fairness, and recommendations?
\item \emph{RQ3.} How can these findings be translated into traceable, provisional early-stage AI design constraints for the proposed tool?
\end{itemize}

The paper makes the following three contributions:

\begin{itemize}
\item \textbf{Exploratory empirical findings on resident perspectives on AI-driven wildfire risk assessment.} 

These findings integrate survey measures with qualitative town-hall discussion in Los Angeles County and identify resident concerns related to privacy, fairness, uncertainty, explanation, and mitigation feasibility.

\item \textbf{The AI Value Map, a structured traceability artifact for early-stage community-grounded AI design.} 

The Map organizes participant evidence about the proposed assessment, its anticipated outputs and use, and practical conditions affecting action into value dimensions, provisional design requirements, and possible consequences if those requirements are not addressed.

\item \textbf{A pre-model mapping procedure for AI hazard-tool design.} 

The procedure shows how early community findings can be organized before a predictive model or functional interface is built and may be adapted to other hazard settings such as flood, heat, or air quality.
\end{itemize}

These contributions position the AI Value Map as a practical, pre-model method for translating community evidence into design constraints, not as a new theory of values in design.
The rest of the paper is organized as follows. Section \ref{sec:background} reviews literature on wildfire risk communication, AI-based wildfire assessment, explainable AI, and participatory design. Section \ref{sec:study_framing} presents the study's pre-model community value alignment framing and case study selection, and Section \ref{sec:methods} describes the mixed-methods design. Section \ref{sec:findings} reports the empirical integrated findings, Section \ref{sec:ai_value_map} presents the AI Value Map, and Section \ref{sec:discussion} discusses implications for AI hazard-tool design. Section \ref{sec:future_work} outlines the study's limitations and future work, and Section \ref{sec:conclusion} concludes the paper.

%% file: sections/02_background.tex
\section{Background and Research Gap}
\label{sec:background}

The proposed tool sits at the intersection of four bodies of literature: wildfire risk communication, AI-based wildfire risk assessment, explainable AI for non-experts, and participatory and value-sensitive design. Each is reviewed briefly below, ending with the gap that motivates the AI Value Map.


\subsection{Wildfire risk communication and the equity dimension}

Wildfire risk is closely related to the expansion of the wildland--urban interface (WUI), where structures and other human development meet or intertwine with wildland vegetation \citep{USFireAdministrationND}. As development expands into these areas, the exposure to wildfires is shaped by the interaction of vegetation, topography, neighborhood patterns, building design, access, and exits \citep{AmericanPlanningAssociation2016}. Planning and engineering research identifies these factors as important for community wildfire resilience and also points to inconsistencies between planning, construction, and governance approaches in local governments \citep{GonzalezMathiesenMarch2018,Penney2025,Penney2024}.

Wildfire research has also examined how residents understand risk and decide whether to take protective action. The \emph{Protective Action Decision Model} describes these decisions as shaped by environmental and social cues, perceived threat, perceptions of available protective actions, and views of the information sources involved \citep{LindellPerry2012}. In wildfire settings, prior experience, trust, perceived ability to act, and local conditions can also influence preparedness and mitigation decisions \citep{McCaffrey2006,SteelmanMcCaffrey2013,Stidham2014}. Research on wildfire communication therefore emphasizes information that is clear, specific, timely, and connected to local conditions \citep{McCaffrey2006,SteelmanMcCaffrey2013}.

Wildfire vulnerability is also unevenly distributed across communities. Majority-Black, Hispanic, and Native American census tracts have been found to experience greater wildfire vulnerability than other tracts because of differences in both wildfire exposure and adaptive capacity \citep{Davies2018}. Research on California wildfires has similarly documented how emergency response and recovery systems may overlook undocumented Latino/a and Indigenous migrant communities \citep{MendezFloresHaroZucker2020}.


\subsection{Wildfire risk prediction advances and the resident comprehension gap}

Machine-learning applications in wildfire science have expanded across fire detection and mapping, wildfire occurrence and risk assessment, fire-behavior prediction, and management support \citep{Jain2020}. Recent work has also developed remote-sensing datasets and benchmark models for predicting wildfire spread one day ahead \citep{Huot2022}. These advances expand the technical capacity available for estimating and mapping wildfire risk, but predictive performance alone does not determine whether a resident-facing tool will be understandable or useful. Explainable-AI and ethical-AI research emphasizes the importance of interpretable outputs, transparency, fairness, and appropriate user trust \citep{Ribeiro2016,Floridi2018,Liao2020}. However, the literature reviewed here provides limited empirical guidance on how wildfire risk information should be presented to residents or how resident needs should shape explanations and other design decisions before a model is developed.


\subsection{Explainable AI and the limits of technical interpretability for residents}

Explainable AI (XAI) offers methods and toolkits for examining model behavior and individual predictions, including AI Explainability 360 \citep{Arya2019}, the What-If Tool \citep{Wexler2019}, LIME \citep{Ribeiro2016}, and SHAP \citep{LundbergLee2017}. These resources support local explanation, feature attribution, hypothetical testing, model comparison, and analysis across data subsets. SHAP-based methods have also been applied to wildfire susceptibility mapping \citep{Iban2024}. However, technical interpretability does not determine by itself how risk information should be explained to residents. User-centered XAI and human--AI interaction research emphasizes designing explanations around users' questions and contexts and clearly communicating system capabilities and limitations \citep{Liao2020,Amershi2019}. Ethical and human-centered AI frameworks provide broader guidance on fairness, transparency, reliability, oversight, and trustworthiness \citep{Shneiderman2020,Floridi2018}. These works provide important foundations, but they do not offer a case-specific traceability procedure showing how early resident evidence can be translated into design requirements for a proposed hazard tool.


\subsection{Participatory and value-sensitive design for AI tools}

Value Sensitive Design (VSD) provides a structured approach for considering human values through conceptual, empirical, and technical investigations \citep{FriedmanHendry2019}. It has also been adapted to AI development to connect ethical values with specific design requirements \citep{UmbrelloVandePoel2021}. Co-design and participatory design similarly involve intended users in shaping technology, not treating them only as sources of feedback \citep{SandersStappers2008}. Participatory AI research, however, warns that participation can become largely symbolic when communities are consulted without meaningful influence over important design decisions \citep{Sloane2022,Birhane2022}. Therefore, for this study, the key question is  not simply whether residents are included, but whether they are involved early enough to shape decisions about data, explanations, privacy, fairness, and recommendations.


\subsection{The integrated gap}

These four bodies of literature show why resident-facing AI hazard tools need to be understandable, trustworthy, and responsive to local conditions. Wildfire research has examined how residents understand risk and decide whether to take protective action. AI and XAI research provide methods for modeling risk and examining model outputs, but many focus on technical prediction or interpretability rather than how risk information should be communicated to residents. Participatory design and Value Sensitive Design provide established ways to identify stakeholder values and involve communities in technology design. What remains less clear is how a research team can document the path from early community evidence to specific design requirements before key decisions about the AI model and interface are unchangeable. The next section explains how this study was framed around this gap.

%% file: sections/03_study_framing.tex
\section{Study Framing and Case Setting}
\label{sec:study_framing}

This study was designed around three methodological commitments. First, residents' values, concerns, and practical expectations related to the proposed AI assessment and its anticipated outputs should be examined while core design choices are still open. Second, this evidence should be grounded in a specific proposed tool, geographic context, and risk exposure and not be treated in abstract. Third, the link between resident input and design decisions should be made explicit through a structured artifact, rather than left to informal interpretation by designers or researchers.


\subsection{Pre-model engagement as a methodological commitment}

In many AI-development processes, community engagement occurs after a model or prototype has already been developed. At this stage, engagement can help identify usability problems and improve how explanations are presented, but it may have limited ability to change deeper design choices, such as what the tool does, what data it collects, how it works, and what information it provides to users. For AI hazard tools meant to serve communities with unequal wildfire exposure and unequal ability to prepare or recover, this late-stage approach may not be sufficient \citep{Birhane2022}. Several important design decisions may already be largely fixed before community engagement begins. Engagement can still help refine how those decisions are implemented, but it may have limited influence on the underlying choices.

We conducted this study before reaching that stage. Data collection ran from December 2025 through mid-March 2026, and throughout this period the proposed AI-driven wildfire risk-assessment tool existed only as wireframes, scenario materials, and conceptual flows. No predictive model had been trained, no model-generated risk scores had been produced, and the interface remained conceptual. Participants therefore responded not to a functioning AI system, but to a proposed resident-facing tool and simulated examples of the outputs it might eventually provide. This distinction is important for interpreting the findings. The findings in Section \ref{sec:findings} do not describe how residents use or interpret an existing AI tool; instead, they show how residents responded to conceptual materials at a point when their concerns could still affect design requirements.


\subsection{Proposed AI-driven wildfire risk assessment tool}

The proposed tool is a resident-facing smartphone application for parcel-level wildfire risk assessment. Its purpose is to give residents in fire-prone areas a more direct and accessible way to understand wildfire risk conditions around their own properties. Unlike large-scale risk maps or expert inspection reports, the tool is intended to provide property-specific feedback that residents can interpret and use in everyday mitigation decisions.

\begin{figure}[ht]
\centering
\includegraphics[width=0.90\textwidth]{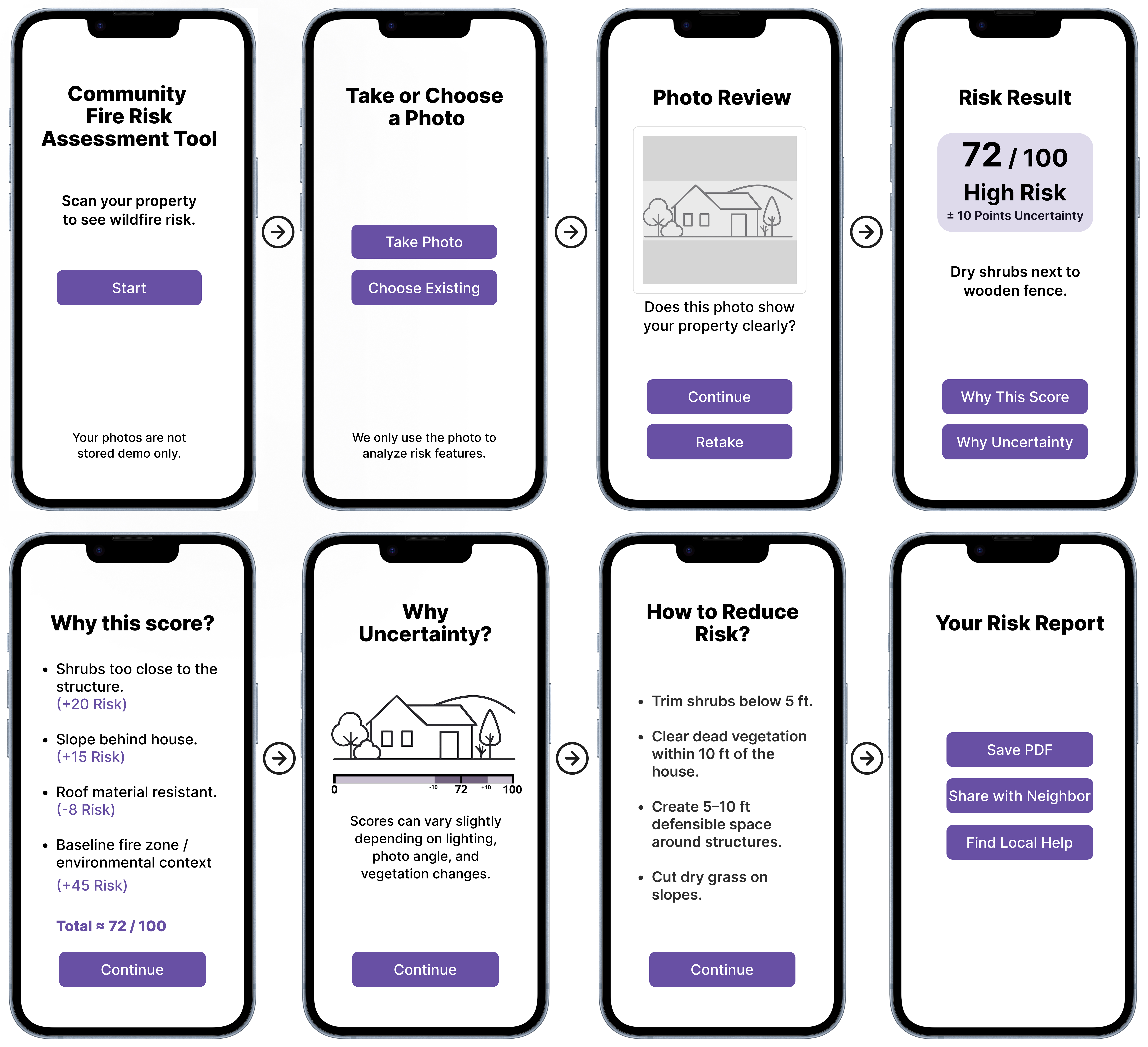}
\caption{Eight-screen conceptual wireframe sequence of the proposed AI-driven wildfire risk assessment tool, as presented to survey and town hall participants. The 72/100 risk score, ±10 points uncertainty, and \emph{“Why this score / Why uncertainty / How to Reduce Risk”} panels are illustrative; no predictive model had been built at the time of the study. The \emph{“Your Risk Report”} screen supports saving and sharing the assessment result. Source: Authors’ own work.}\label{fig2}
\end{figure}

In the proposed workflow for this study, wireframes presented an eight-screen user journey. Users would begin by taking or uploading recent exterior images of their property and then reviewing the image to confirm that it showed the property clearly. They would then receive a simulated risk result, illustrated in Fig. \ref{fig2} as a score of 72/100, classified as high risk, with a $\pm$10-point uncertainty range. The result screen also provided a short summary of a main visible risk factor. Supporting screens explained the result in more detail. The \emph{``Why this score?''} screen presented contributors to the score, including nearby shrubs, slope, roof material, and fire-zone context. The \emph{``Why uncertainty?''} screen explained that scores could vary depending on factors such as lighting and photo angle. The \emph{``How to reduce risk?''} screen provided suggested actions, including trimming shrubs, clearing dead vegetation, and creating defensible space. A final \emph{``Your Risk Report''} screen allowed users to save the report as a PDF, share it with a neighbor, or find local assistance.

The score, explanation factors, uncertainty information, and mitigation recommendations shown in the wireframes were illustrative outputs used to communicate the proposed concept. Participants were asked to respond to the workflow and outputs as a design concept, including whether they found the information understandable, trustworthy, and useful for taking action. Figure \ref{fig2} presents the wireframe sequence used in the study.

\subsection{Case selection: Los Angeles County, CA}

We selected Los Angeles County as the case context for this study because it combines recurring wildfire exposure with the social and housing diversity that makes resident-facing risk communication especially important. The county includes extensive wildland--urban interface areas, including places designated by California Department of Forestry and Fire Protection (CAL FIRE) as Very High Fire Hazard Severity Zones, and contains substantial demographic, economic, and housing variation that makes equity a central concern in hazard-tool design \citep{Penney2025, MendezFloresHaroZucker2020}. Across the county, drought, heat, vegetation conditions, and continued development in fire-prone areas contribute to recurring wildfire risk. The January 2025 Palisades and Eaton Fires made these issues especially visible by affecting communities with different landscape conditions, housing patterns, and recovery challenges \citep{CALFIREEaton2025, CALFIREPalisades2025}.

Pacific Palisades, Altadena, and Pasadena illustrate this variation within the county. Pacific Palisades includes coastal hillside neighborhoods with steep slopes, dense vegetation, and high-value housing. Altadena sits in the foothills below the San Gabriel Mountains and has a long-established residential character shaped in part by African American settlement and middle-class home ownership \citep{Ong2025}. Pasadena includes mixed residential and commercial areas and a more varied demographic profile. The Eaton Fire also drew attention to disparities in damage within Altadena, where historically Black neighborhoods were among the affected areas \citep{Ong2025}. 

These communities shaped the study context and informed the design issues considered for the proposed tool, particularly neighborhood-scale risk, housing variation, fairness, and differences in recovery capacity. Recruitment was conducted across Los Angeles County, and the study did not compare individual communities. This county-wide scope also highlights the need for a resident-facing tool that can accommodate different housing situations, property types, and levels of authority to make property changes.

%% file: sections/04_methods.tex
\section{Research Methods}
\label{sec:methods}


\subsection{Research design}

In this study we used a convergent mixed-methods design \citep{CreswellPlanoClark2018}. We collected two complementary forms of data: structured survey responses from Los Angeles County residents and discussion data from a live virtual town hall meeting. The survey provided patterned evidence about how residents responded to the proposed AI wildfire-risk tool, while the town hall discussion helped us explain the reasoning, values, and concerns behind those responses. The two data types were analyzed separately and then brought together during interpretation. We used one shared engagement protocol so that participants responded to the same core tool concept and visual materials. The final dataset included a structured online survey completed by 30 Los Angeles County residents, including seven who participated in a synchronous virtual town hall meeting. The study was approved by the University of North Carolina at Charlotte Institutional Review Board (IRB-26-0207; approved November 21, 2025). 


\subsection{Recruitment and participants}

For recruitment we used a targeted convenience sampling strategy \citep{Sedgwick2013} directed toward adult residents of Los Angeles County, including homeowners and renters. Recruitment materials were distributed through community networks, social media, partner channels, and printed flyers. Participants were offered a \$20 electronic Amazon gift card after completing the survey. In the recruitment materials we specifically encouraged participation from residents of Altadena, Pasadena, and Pacific Palisades because these communities informed the wildfire context of the proposed tool; however, eligibility was open to residents across Los Angeles County. Although the most severe property damage from the January 2025 fires occurred in and around these communities, their effects extended across the county through widespread smoke, degraded air quality, and broader public awareness of the consequences of severe wildfires. County-wide recruitment therefore allowed the study to include residents who may not have experienced direct property damage but were still familiar with the fires and their wider effects. In the final sample, 10.0\% of respondents (3 of 30) reported living in Pasadena, and none reported living in Altadena or Pacific Palisades; the remaining 90.0\% lived elsewhere in Los Angeles County. This distribution is reported to describe the sample, and no neighborhood-level comparisons were conducted.


\subsection{Sample size justification}

The sample size was chosen to support exploratory mixed-methods analysis. Sample-size adequacy in mixed-methods research depends on the analytical purpose of the study and the kinds of claims being made \citep{OnwuegbuzieCollins2007, Sandelowski1995}. At $n = 30$, the survey supports descriptive summaries and exploratory association analysis; the study does not attempt population-level inference. For the qualitative components, including the seven-participant town hall and open-ended survey responses, prior work on people responding to interface materials indicates that 5--15 participants can surface many comprehension issues \citep{Nielsen1994, Virzi1992}. We used these components to find design-relevant concerns, not to claim representativeness or saturation on themes. Statistical analyses are therefore interpreted as indicative within-sample patterns rather than definitive estimates. Limitations related to sample size and recruitment are addressed in Section \ref{sec:future_work}.


\subsection{Data collection and town hall procedure}

Data collection began with a 90-minute virtual community town hall meeting held via Zoom on December 10, 2025. The session introduced the local wildfire-risk context, the proposed visual assessment concept and workflow, scenario cards showing wildfire-risk factors, and future conceptual app wireframes. Survey sections were placed between presentation and facilitated discussion segments so that participants responded after reviewing the related materials.

The live town hall followed a fixed thirteen-segment agenda, detailed in Table \ref{tab:town-hall-agenda}. It included four facilitated discussion segments and three online Qualtrics survey segments. In the discussion prompts we asked participants about their wildfire-risk experiences, reactions to the visual assessment workflow, missing risk factors they thought the tool should consider, and expectations for a future app. Participants responded verbally or through Zoom chat. With their consent, the discussion segments were recorded and transcribed for thematic analysis.

After the live event, the same online Qualtrics survey link was distributed with access to the recorded material presentation session, allowing additional respondents to review the study materials before completing the survey \citep{Hosseini2026}. The survey remained open until mid-March 2026. The seven town hall participants completed the survey during the live session, while the remaining 23 respondents completed it after viewing the recorded materials.


\subsection{Survey instrument}

The survey was developed for this study and administered anonymously through UNC Charlotte's secure Qualtrics platform. The questions were written to be understandable for non-expert participants and to support the exploratory analyses used in the study. Item design was informed by guidance on Likert-scale construction \citep{Jebb2021}, and questionnaire wording \citep{KrosnickPresser2010}. The research team reviewed the instrument before IRB submission.

The final survey included 31 questions organized into five sections: Background and Household Context (Q1--Q4); Wildfire Experience, Risk Perception, and Preparedness (Q5--Q16); Visual Assessment: Trust, Comfort, and Privacy (Q17--Q23); Feedback on Conceptual App Screens (Q24--Q28); and Final Reflections and App Adoption (Q29--Q31). The survey included binary, nominal, ordinal, five-point Likert, multi-select, and open-ended questions. Five-point Likert items were used for the main attitudinal measures so that responses could be summarized descriptively and examined through exploratory association analyses. Table \ref{tab:survey-instrument} summarizes the survey items, coded variables, response formats, and measurement types used in the analysis.

\begingroup
\footnotesize

\makeatletter
\setlength{\LTcapwidth}{\textwidth}
\def\LT@makecaption#1#2#3{%
  \LT@mcol\LT@cols l{%
    \parbox[t]{\LTcapwidth}{%
      \normalfont\small
      \sbox\@tempboxa{#1{\textbf{#2}\quad}#3}%
      \ifdim\wd\@tempboxa>\hsize
        #1{\textbf{#2}\quad}#3%
      \else
        \box\@tempboxa
      \fi
      \endgraf\vskip\baselineskip}%
  }}%
\makeatother

\setlength{\tabcolsep}{2pt}
\setlength{\LTleft}{0pt}
\setlength{\LTright}{0pt}
\renewcommand{\arraystretch}{1.3}

\begin{longtable}{@{}
>{\centering\arraybackslash}p{0.08\textwidth}
>{\raggedright\arraybackslash}p{0.17\textwidth}
>{\centering\arraybackslash}p{0.10\textwidth}
>{\raggedright\arraybackslash}p{0.135\textwidth}
>{\raggedright\arraybackslash}p{0.465\textwidth}
@{}}

\caption{Town hall agenda and discussion prompts (90 minutes)}
\label{tab:town-hall-agenda}\\[3pt]

\toprule
\textbf{Order} &
\textbf{Segment} &
\textbf{Approx. duration} &
\textbf{Activity type} &
\textbf{Discussion prompt or content focus} \\
\midrule
\endfirsthead

\toprule
\textbf{Order} &
\textbf{Segment} &
\textbf{Approx. duration} &
\textbf{Activity type} &
\textbf{Discussion prompt or content focus} \\
\midrule
\endhead

\endfoot

\botrule
\multicolumn{5}{@{}p{\dimexpr\textwidth-2\tabcolsep\relax}@{}}{%
\textit{Note:} total time = 90 minutes. Durations are approximate and were adjusted in real time to accommodate participant engagement.
}\\
\endlastfoot

1 &
Welcome \& opening remarks &
5 min &
Introduction &
Study purpose, consent, IRB context \\

2 &
Survey Part 1 &
10 min &
Survey (Sections 1--2 of instrument) &
Background \& Household Context; Wildfire Experience, Risk Perception \& Preparedness  \\

3 &
Understanding local wildfire risk &
8 min &
Presentation &
LA County WUI exposure; demographic vulnerability; documented damage patterns from Eaton and Palisades fires \\

4 &
Discussion 1 --- Fire behavior \& lived experience &
8 min &
Facilitated discussion &
``In the fire examples we had, what fire behaviors like spread direction or how close it came to homes, look familiar or concerning to you?'' \\

\midrule

5 &
Solution: How visual assessment could help &
5 min &
Presentation &
Five-step conceptual flow (Input $\rightarrow$ Feature Detection $\rightarrow$ Model Assessment $\rightarrow$ Output $\rightarrow$ Recommendations) \\

6 &
Discussion 2 --- Reactions to visual assessment flow &
8 min &
Facilitated discussion &
``Looking at this visual assessment flow, what part stood out to you the most? Why?'' \\

\midrule

7 &
Scenario cards \& fire risk factors &
5 min &
Presentation &
Four scenario cards (Shrubs + Wood Fence; Overgrown Yard + Close Houses; Cleared Yard + Resistant Roof; Sloped Property with Dry Grass); visual assessment categories (Vegetation, Terrain, Neighborhood Layout, Roof \& Yard Structures) \\

8 &
Discussion 3 --- Scenario cards \& missing elements &
8 min &
Facilitated discussion &
``We've looked at several external factors around homes that affect wildfire risk, now is there another element or issue you have seen that we have not covered?'' \\

\midrule

9 &
Survey Part 2 &
8 min &
Survey (Section 3 of instrument) &
Visual Assessment: Trust, Comfort \& Privacy \\

10 &
Future app wireframes &
5 min &
Presentation &
Eight-screen wireframe sequence (Landing $\rightarrow$ Photo capture $\rightarrow$ Photo review $\rightarrow$ Risk result with 72/100 example $\rightarrow$ Why this score $\rightarrow$ Why uncertainty $\rightarrow$ How to reduce risk $\rightarrow$ Your risk report) \\

11 &
Discussion 4 --- How a future app might look &
8 min &
Facilitated discussion &
``How a future app might look --- reactions to the wireframe sequence and what features would feel useful or missing in a final version of the app?'' \\

\midrule

12 &
Survey Part 3 &
8 min &
Survey (Sections 4--5 of instrument) &
Feedback on Conceptual App Screens; Final Reflections \& App Adoption \\

13 &
Closing \& final remarks &
4 min &
Closing &
Thank-you; gift card distribution instructions \\

\end{longtable}
\endgroup


\subsection{Sample demographics}

The collected sample reflected working-age, mostly homeowner LA County residents with high smartphone access. Table \ref{tab:sample-demographics} presents the demographic distribution. The sample was unevenly distributed among 30-44-year-olds (46.7\%), homeowners (63.3\%) and single-family home occupants (56.7\%). Technology access was near-universal (96.7\%). Insurance uncertainty was notable; as although 80.0\% reported property insurance, 30.0\% could not confirm whether their policy covered wildfire damage. This uncertainty is relevant to interpreting adoption-related responses and practical barriers in Section \ref{sec:findings}. Table \ref{tab:sample-demographics} presents the detailed demographic characteristics of the participants.


\subsection{Data preparation}

Survey data exported from Qualtrics were screened for consent, completeness, and response consistency before analysis. All 30 respondents provided consent and were retained in the analytic sample. Most survey items were complete. Among the Likert and ordinal items, Q11 and Q12 each had one missing response. Among the multi-select items, Q15 and Q21 each had one blank response, and Q22 had two blank responses. Q5a was answered by five respondents because it was shown only to participants who reported prior wildfire experience in Q5. Missing values were retained as missing rather than imputed. 

For multi-select items, percentages were calculated using the number of respondents who answered each item; blank responses were excluded from the denominator. Categorical variables were coded as nominal or binary, five-point Likert items were retained as ordinal, and selected options in multi-select responses were converted into binary indicator variables. Open-ended responses from Q25 and Q31, along with transcripts from the four town hall discussion segments, were prepared separately for thematic analysis.

\begingroup
\footnotesize

\makeatletter
\setlength{\LTcapwidth}{\textwidth}
\def\LT@makecaption#1#2#3{%
  \LT@mcol\LT@cols l{%
    \parbox[t]{\LTcapwidth}{%
      \normalfont\small
      \sbox\@tempboxa{#1{\textbf{#2}\quad}#3}%
      \ifdim\wd\@tempboxa>\hsize
        #1{\textbf{#2}\quad}#3%
      \else
        \box\@tempboxa
      \fi
      \endgraf\vskip\baselineskip}%
  }}%
\makeatother

\setlength{\tabcolsep}{2pt}
\setlength{\LTleft}{0pt}
\setlength{\LTright}{0pt}
\renewcommand{\arraystretch}{1.2}

\begin{longtable}{@{\extracolsep{\fill}}
>{\raggedright\arraybackslash}p{0.15\textwidth}
>{\centering\arraybackslash}p{0.04\textwidth}
>{\raggedright\arraybackslash}p{0.205\textwidth}
>{\raggedright\arraybackslash}p{0.20\textwidth}
>{\raggedright\arraybackslash}p{0.205\textwidth}
>{\raggedright\arraybackslash}p{0.12\textwidth}
@{}}

\caption{Summary of survey instrument}\label{tab:survey-instrument}\\[3pt]

\toprule
\textbf{Section} &
\textbf{Q\#} &
\textbf{Question (abbreviated)} &
\textbf{Coded variable} &
\textbf{Response format} &
\textbf{Variable type} \\
\midrule
\endfirsthead

\toprule
\textbf{Section} &
\textbf{Q\#} &
\textbf{Question (abbreviated)} &
\textbf{Coded variable} &
\textbf{Response format} &
\textbf{Variable type} \\
\midrule
\endhead

\endfoot

\botrule
\multicolumn{6}{@{}p{\dimexpr\textwidth-2\tabcolsep\relax}@{}}{%
\textit{Note:} Variable names follow conventions used throughout the later analyses in Sections \ref{sec:findings} and \ref{sec:ai_value_map}.
}\\
\endlastfoot

\multirow[t]{4}{=}{\textbf{1. Background \& Household Context}}
& Q1
& Age group
& Age
& 18--29, 30--44, 45--59, 60--74, 75+
& Ordinal \\

& Q2
& Area of residence
& Area
& Pacific Palisades / Pasadena / Altadena / None of the above
& Nominal (multi-select) \\

& Q3
& Home tenure
& Tenure
& Owned / Rented / Other
& Nominal \\

& Q4
& Home type
& Home\_\allowbreak Type
& Single-family / Townhouse / Condo / Apartment / Other
& Nominal \\

\midrule

\multirow[t]{12}{=}{\textbf{2. Wildfire Experience, Risk Perception \& Preparedness}}
& Q5
& Prior wildfire evacuation or damage
& Prior\_\allowbreak Wildfire
& Yes / No
& Binary \\

& Q5a
& Type of impact
& Impact\_\allowbreak Type
& Evacuation / Damage / Power shutoff / Smoke / Other
& Nominal (multi-select) \\

& Q6
& Personally affected by Palisades/Eaton/Hughes fires
& Recent\_\allowbreak Fires
& Yes / No / Not sure
& Nominal \\

& Q7
& Has homeowners/renters insurance
& Insurance
& Yes / No / Not sure
& Nominal \\

& Q8
& Insurance includes wildfire coverage
& Wildfire\_\allowbreak Coverage
& Yes / No / Not sure
& Nominal \\

& Q9
& Pre-occupancy fire-safety inspection
& Pre\_\allowbreak Inspection
& Yes / No / Not sure
& Nominal \\

& Q10
& Concern about neighborhood wildfire risk
& Concern
& 5-level Likert (Not at all $\rightarrow$ Extremely concerned)
& Ordinal \\

& Q11
& Perceived neighborhood risk
& Perceived\_\allowbreak Risk
& 5-level Likert
& Ordinal \\

& Q12
& Priority of wildfire safety for household
& Priority
& 5-level Likert
& Ordinal \\

& Q13
& Prevention steps taken
& Prevention\_\allowbreak Steps
& Multi-select list
& Nominal (multi-select) \\

& Q14
& Current information sources
& Info\_\allowbreak Sources
& Multi-select list
& Nominal (multi-select) \\

& Q15
& Apps/websites/alert systems used
& Tools\_\allowbreak Used
& Multi-select list + open-ended
& Nominal (multi-select) \\

& Q16
& Primary smartphone/device
& Device
& iPhone / Android / Other / None
& Nominal \\

\midrule

\multirow[t]{7}{=}{\textbf{3. Visual Assessment: Trust, Comfort \& Privacy}}
& Q17
& Comfort with visual-assessment tool
& Comfort\_\allowbreak VisualAssessment
& 5-level Likert
& Ordinal \\

& Q18
& Usefulness of fire-score metric
& Score\_\allowbreak Useful
& 5-level Likert agreement
& Ordinal \\

& Q19
& Usefulness of personalized recommendations
& Reco\_\allowbreak Useful
& 5-level Likert
& Ordinal \\

& Q20
& Willingness to take action from app insights
& Willingness\_\allowbreak Act
& 5-level Likert
& Ordinal \\

& Q21
& Barriers to action
& Action\_\allowbreak Barriers
& Multi-select list + open-ended
& Nominal (multi-select) \\

& Q22
& Concerns about app use
& App\_\allowbreak Concerns
& Multi-select list + open-ended
& Nominal (multi-select) \\

& Q23
& Perceived fairness across neighborhoods
& Fairness
& 5-level Likert
& Ordinal \\

\midrule

\multirow[t]{5}{=}{\textbf{4. Feedback on Conceptual App Screens}}
& Q24
& Reaction to overall app layout
& Layout\_\allowbreak Reaction
& 5-level Likert
& Ordinal \\

& Q25
& What is unclear or missing in screens
& Missing\_\allowbreak Features
& Open-ended
& Qualitative \\

& Q26
& Preferred risk-score format
& Format\_\allowbreak Preference
& Number / Color band / Descriptive label
& Nominal \\

& Q27
& Likelihood of using Save/Share button
& SaveShare\_\allowbreak Likelihood
& 5-level Likert
& Ordinal \\

& Q28
& Recipients of shared result
& Share\_\allowbreak With
& Multi-select list
& Nominal (multi-select) \\

\midrule

\multirow[t]{3}{=}{\textbf{5. Final Reflections \& App Adoption}}
& Q29
& Likelihood of using app if available
& Adoption\_\allowbreak Likelihood
& 5-level Likert
& Ordinal \\

& Q30
& Willingness to share result for community safety
& Community\_\allowbreak Sharing
& 5-level Likert
& Ordinal \\

& Q31
& What would encourage regular use
& Adoption\_\allowbreak Drivers
& Open-ended
& Qualitative \\

\end{longtable}
\endgroup


\subsection{Quantitative analysis}

Quantitative analysis included descriptive summaries and exploratory Spearman rank-correlation analysis. Analyses were conducted in Python using pandas, NumPy, SciPy, and statsmodels, with seaborn and matplotlib used for visualization. Descriptive statistics were used to summarize participant characteristics, response frequencies, and Likert-scale items. For ordinal Likert items, means and standard deviations were used as descriptive summaries.

We used Spearman rank-order correlation analysis to examine whether higher responses on one ordinal survey item tended to occur with higher responses on another. Spearman correlation was selected instead of Pearson correlation because the survey items used ordinal five-point response scales and may not meet parametric assumptions \citep{deWinter2016}. The correlation matrix included twelve items: Q10, Q11, Q12, Q17, Q18, Q19, Q20, Q23, Q24, Q27, Q29, and Q30. All 66 unique pairwise combinations among these items were evaluated. Missing responses were removed separately for each pair, resulting in pairwise sample sizes from $n = 28$ to $n = 30$.

Spearman's $\rho$ and two-sided asymptotic $p$-values were calculated using \emph{scipy.stats.spearmanr}. Because all 66 unique pairwise associations were examined, we applied the Benjamini--Hochberg procedure using \emph{statsmodels.stats.multitest.multipletests} to control the false discovery rate (FDR). Twenty-one associations remained significant at $p_{\mathrm{FDR}} < .05$. Section \ref{sec:findings} reports 10 of these associations because they were directly relevant to the integrated interpretation. Because the study used a small, non-probability sample, all correlations and associated $p$-values are interpreted cautiously as exploratory within-sample patterns and not as population-level evidence.

\begin{table}[h]
\centering
\caption{Sample demographics ($n = 30$)}
\label{tab:sample-demographics}

\footnotesize
\setlength{\tabcolsep}{3pt}
\renewcommand{\arraystretch}{1.20}

\begin{tabularx}{0.80\linewidth}{@{}
>{\raggedright\arraybackslash}p{0.20\linewidth}
>{\raggedright\arraybackslash}X
>{\centering\arraybackslash}p{0.10\linewidth}
>{\centering\arraybackslash}p{0.14\linewidth}
@{}}

\toprule
\textbf{Variable} &
\textbf{Category} &
\textbf{Frequency} &
\textbf{Proportion (\%)} \\
\midrule

\multirow[t]{5}{=}{Age (Q1)}
& 18--29 & 7 & 23.3 \\
& 30--44 & 14 & 46.7 \\
& 45--59 & 6 & 20.0 \\
& 60--74 & 2 & 6.7 \\
& 75+ & 1 & 3.3 \\

\midrule

\multirow[t]{4}{=}{Area of residence (Q2)}
& Pasadena & 3 & 10.0 \\
& Altadena & 0 & 0.0 \\
& Pacific Palisades & 0 & 0.0 \\
& None of the above (other LA County) & 27 & 90.0 \\

\midrule

\multirow[t]{3}{=}{Home tenure (Q3)}
& Owned by self/family & 19 & 63.3 \\
& Rented & 9 & 30.0 \\
& Other & 2 & 6.7 \\

\midrule

\multirow[t]{5}{=}{Home type (Q4)}
& Single-family house & 17 & 56.7 \\
& Apartment unit & 6 & 20.0 \\
& Condominium & 5 & 16.7 \\
& Townhouse & 1 & 3.3 \\
& Other & 1 & 3.3 \\

\midrule

\multirow[t]{3}{=}{Primary device (Q16)}
& iPhone & 23 & 76.7 \\
& Android & 6 & 20.0 \\
& No smartphone & 1 & 3.3 \\

\midrule

\multirow[t]{3}{=}{Insurance coverage (Q7)}
& Homeowners/renters insurance & 24 & 80.0 \\
& No insurance & 1 & 3.3 \\
& Not sure & 5 & 16.7 \\

\midrule

\multirow[t]{3}{=}{Wildfire-specific coverage (Q8)}
& Yes & 19 & 63.3 \\
& No & 2 & 6.7 \\
& Not sure & 9 & 30.0 \\

\bottomrule
\end{tabularx}
\end{table}


\subsection{Qualitative analysis}

We analyzed the qualitative data using inductive thematic analysis \citep{BraunClarke2008}. The data included open-ended responses to Q25 ($n = 22$) and Q31 ($n = 24$), after excluding non-substantive entries such as ``N/A,'' as well as transcripts from four town hall discussion segments. The analysis focused on participants' expressed meanings, concerns, and expectations related to the proposed tool. The first author reviewed the survey comments and town hall transcripts in several rounds. Initial codes were assigned to relevant excerpts and then grouped into candidate themes based on recurring ideas and design-related concerns. The candidate themes were checked against the original excerpts to confirm that they were supported by the data and sufficiently distinct from one another. Codes and theme boundaries were revised during this process, and comments that did not fit the emerging themes were retained and considered during interpretation. Themes from the survey and town hall data were compared to identify areas of convergence and divergence. Representative quotations were checked against the original responses and transcripts to preserve participants' wording and context. They were used in Section \ref{sec:findings} to illustrate and support the reported findings. 


\subsection{Mixed-methods integration}

Quantitative and qualitative findings were integrated during interpretation using a convergent mixed-methods approach \citep{CreswellPlanoClark2018, Fetters2013}. Integration occurred at two levels. First, descriptive results and exploratory correlation findings were compared with themes from the open-ended survey responses and town hall discussion. Second, the combined findings were used to develop the nine value dimensions of the AI Value Map presented in Section \ref{sec:ai_value_map}. When the survey results and qualitative feedback pointed to similar conclusions, they were treated as reinforcing each other. When they differed, the differences were retained to show the range and complexity of residents' views.

\subsection{Validity, reliability, and ethical considerations}

The research team reviewed the survey before IRB submission and aligned the items with the conceptual framing presented in Sections \ref{sec:introduction}, \ref{sec:background}, and \ref{sec:study_framing}. Reliability and trustworthiness were supported by using the same core study materials across the live town hall and post-event survey, documenting the survey instrument and town hall agenda, and checking qualitative themes against the original survey comments and town hall transcripts.

All survey respondents provided electronic informed consent, and town hall participants verbally reconfirmed their consent before the discussion segments were recorded. Online survey responses were anonymous. The gift-card contact form, which participants could complete at the end of the survey to receive the incentive, was also stored separately so that contact information could not be linked to survey responses. Data were handled in accordance with UNC Charlotte's OneIT Guidelines for Data Handling.

%% file: sections/05_findings.tex
\section{Integrated Findings}
\label{sec:findings}

This section reports the integrated empirical findings from the survey and town hall meeting, with quantitative and qualitative findings presented together in each thematic subsection. The findings are organized around the conditions residents identified as important for designing an AI-driven wildfire risk-assessment tool. As described in Section 4.8, the correlation analyses are exploratory; the reported coefficients and FDR-adjusted $p$-values describe patterns within this sample and should not be generalized to the wider population.


\subsection{Wildfire risk perception and property-level information gap}

Participants reported modest wildfire concern overall, but their concern, assessment of neighborhood risk, household preparedness priorities, and willingness to act were connected within the sample. On the five-point survey scales, participants' average level of concern about wildfire risk in their neighborhood (Q10) was 2.43 ($SD = 0.86$). Their average rating of how at risk they felt their neighborhood was for wildfire (Q11) was 2.41 ($SD = 0.78$), while their average rating of how much of a priority wildfire safety and preparedness was for their household (Q12) was 2.93 ($SD = 0.92$).

Also, based on association analysis, participants who reported greater concern about wildfire in their neighborhood also tended to rate their neighborhood's wildfire risk higher ($Q10 \leftrightarrow Q11$, $n = 29$, $\rho = 0.651$, $p_{\mathrm{FDR}} = .0014$). Higher neighborhood risk ratings were associated with assigning greater priority to wildfire safety and preparedness within the household ($Q11 \leftrightarrow Q12$, $n = 28$, $\rho = 0.598$, $p_{\mathrm{FDR}} = .0037$). Greater priority placed on household preparedness was also associated with stronger willingness to act on guidance provided by the app ($Q12 \leftrightarrow Q20$, $n = 29$, $\rho = 0.513$, $p_{\mathrm{FDR}} = .0147$). 

Many participants were also unsure about wildfire-related conditions at their own properties. One-third (33.3\%) selected \emph{``Don't know''} when asked which wildfire prevention steps had been taken at their property (Q13), 30.0\% could not confirm whether their insurance covered wildfire damage (Q8), and 36.7\% were unsure whether a fire-safety inspection had occurred before they moved in (Q9).

Participants also relied mainly on broad wildfire information sources rather than property-specific tools. The most common sources were social media (80.0\%), news media (76.7\%), and government wildfire websites (43.3\%). Fewer participants reported using neighbors (20.0\%) or the local fire department (20.0\%) as information sources. Among the 29 respondents to Q15, more than half (58.6\%) reported using no wildfire-specific apps, websites, or alert systems; among those who used such tools, “Watch Duty” was named most often. Town hall participants described a similar gap, mentioning that broad media sources were useful but did not always provide the parcel-level details they needed during wildfire events. Several participants said they relied on neighbors for local details that were not available through broader sources.


\subsection{Comfort, usefulness, and willingness to act on visual assessment}

Participants responded more positively to the information the tool would provide than to the proposed visual-assessment process. On the five-point survey scales, participants' average comfort with a wildfire-risk tool that would use visual assessment to analyze property data (Q17) was 3.30 ($SD = 1.12$). This item had the largest standard deviation among the technology-related measures, indicating considerable variation in participants' comfort with the proposed process. Participants' agreement that the wildfire risk score would be useful (Q18) averaged 3.87 ($SD = 0.63$), while their ratings of how useful personalized safety recommendations accompanying the score would be (Q19) averaged 3.80 ($SD = 0.92$). Their average willingness to take action based on insights provided by the app (Q20) was 3.50 ($SD = 0.86$), lower than both usefulness ratings. 

The association analysis also showed that intended app use was related to both comfort with the proposed assessment process and the perceived usefulness of the information the app would provide. Greater comfort with the proposed visual-assessment process was associated with a higher likelihood of using the app if it were available ($Q17 \leftrightarrow Q29$, $n = 30$, $\rho = 0.724$, $p_{\mathrm{FDR}} < .001$). Also, participants who rated the wildfire risk score as more useful were also more likely to say they would use the app ($Q18 \leftrightarrow Q29$, $n = 30$, $\rho = 0.669$, $p_{\mathrm{FDR}} < .001$). Higher ratings of the usefulness of personalized recommendations were associated with both a greater likelihood of using the app ($Q19 \leftrightarrow Q29$, $n = 30$, $\rho = 0.631$, $p_{\mathrm{FDR}} = .0017$) and greater willingness to take action based on its insights ($Q19 \leftrightarrow Q20$, $n = 30$, $\rho = 0.726$, $p_{\mathrm{FDR}} < .001$). 

\begin{figure}[ht]
\centering
\includegraphics[width=\linewidth]{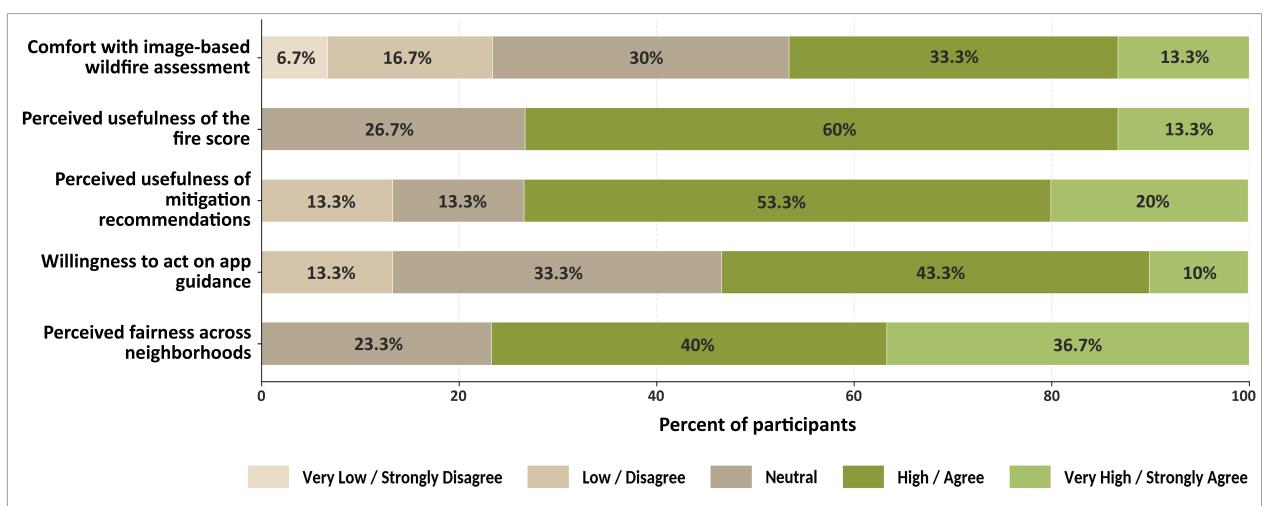}
\caption{Participant perceptions of the visual wildfire risk-assessment system across five attitudinal items: comfort with image-based assessment (Q17), perceived usefulness of the risk score (Q18), perceived usefulness of mitigation recommendations (Q19), willingness to act on app guidance (Q20), and perceived fairness across neighborhoods (Q23). The stacked percentages show that comfort with image-based assessment received the largest shares of neutral and negative responses, while usefulness and fairness received the highest levels of agreement. Willingness to act fell between these patterns. Source: Authors’ own work.}\label{fig3}
\end{figure}

These patterns show that intended use was connected both to comfort with the assessment process and to the perceived usefulness of the score and recommendations. These associations describe patterns within this sample and do not establish causation. Figure \ref{fig3} summarizes the distribution of responses across the five attitudinal items. Comfort with the proposed visual-assessment process received more neutral and negative responses than the items measuring the usefulness of the score, usefulness of the recommendations, and perceived fairness. Fairness across neighborhoods (Q23) is discussed separately in Section 5.4.

The qualitative responses added context to these patterns. Some participants asked, \emph{``What does 72/100 High Risk really mean?''} and wanted the tool to \emph{``highlight areas of the pictures and link them with each risk.''} These comments indicate that comfort depended partly on whether participants could understand the result and connect it to visible property conditions. Other participants emphasized the practical value of \emph{``tracking my property's risk over time,''} \emph{``showing how improvements reduce the score,''} and using the tool to \emph{``save us and others from the wildfires.''} These responses focused on what the tool could help residents do, including monitoring changes over time and identifying practical mitigation steps.

\subsection{Risk-score explanation preferences and feedback on conceptual screens}

When we asked how they preferred the risk score to be presented (Q26), 53.3\% of participants selected a color band, 26.7\% selected a numeric score, and 20.0\% selected a descriptive label. Although color bands were the most common choice, preferences were distributed across all three formats, showing that no single format was preferred by all participants. 

Responses to the conceptual app screens were generally positive. Participants' ratings of the overall look and layout of the app flow (Q24) averaged 4.13 ($SD = 0.86$). This tied with their ratings of how fair the proposed tool would be across different neighborhoods (Q23) as the highest mean among the survey items.

Open-ended Q25 responses, which asked participants what was unclear or missing in the conceptual app screens, identified four recurring concerns. The most common theme concerned how to interpret the risk score. Participants asked \emph{``What does `72/100 High Risk' really mean?''} and suggested combining \emph{``Numbers and different colors to identify the scores.''} A second theme concerned guidance for submitting property photographs, with comments including \emph{``More clear instruction at the beginning''} and \emph{``Please choose a very recent photo.''} A third theme concerned making the system's reasoning visible. Participants wanted the tool to \emph{``highlight areas of the pictures and link them with each risk''} and provide \emph{``colors or score metrics tied to specific features.''} A fourth, smaller theme addressed community-level context, with one participant suggesting \emph{``GPS-based data to show overall risk across the community.''} Several participants reported no concerns, describing the interface as \emph{``Very clear''} or stating, \emph{``None, everything is as clear as it gets,''} which is consistent with the high layout rating.


\subsection{Fairness, privacy, and equity concerns}

Within this sample, participants generally viewed the proposed tool concept as fair across different neighborhoods. Participants' ratings of how fair they thought the proposed tool would be across neighborhoods (Q23) averaged 4.13 ($SD = 0.78$) on the five-point scale, and 76.7\% rated it as somewhat or very fair. Because no working tool existed, these responses reflect participants' initial views of the concept. At the same time, privacy and data security were the most common concerns reported in the survey.

Town hall discussions showed that participants understood fairness as more than applying the same assessment process across neighborhoods. In Discussions 2 and 3, they raised concerns about risk factors the proposed assessment might overlook, including wind speed, eucalyptus and other fire-prone vegetation, neighboring-property maintenance, vacant lots, and human-caused ignition. One participant noted, \emph{``I heard eucalyptus tree catches more fire as it generates oil while burning and cause more damage.''} Based on these comments, participants linked fairness to whether the proposed assessment accounted for relevant local conditions, including conditions that could not be identified from a property photograph alone.

\begin{figure}[ht]
\centering
\includegraphics[width=0.90\textwidth]{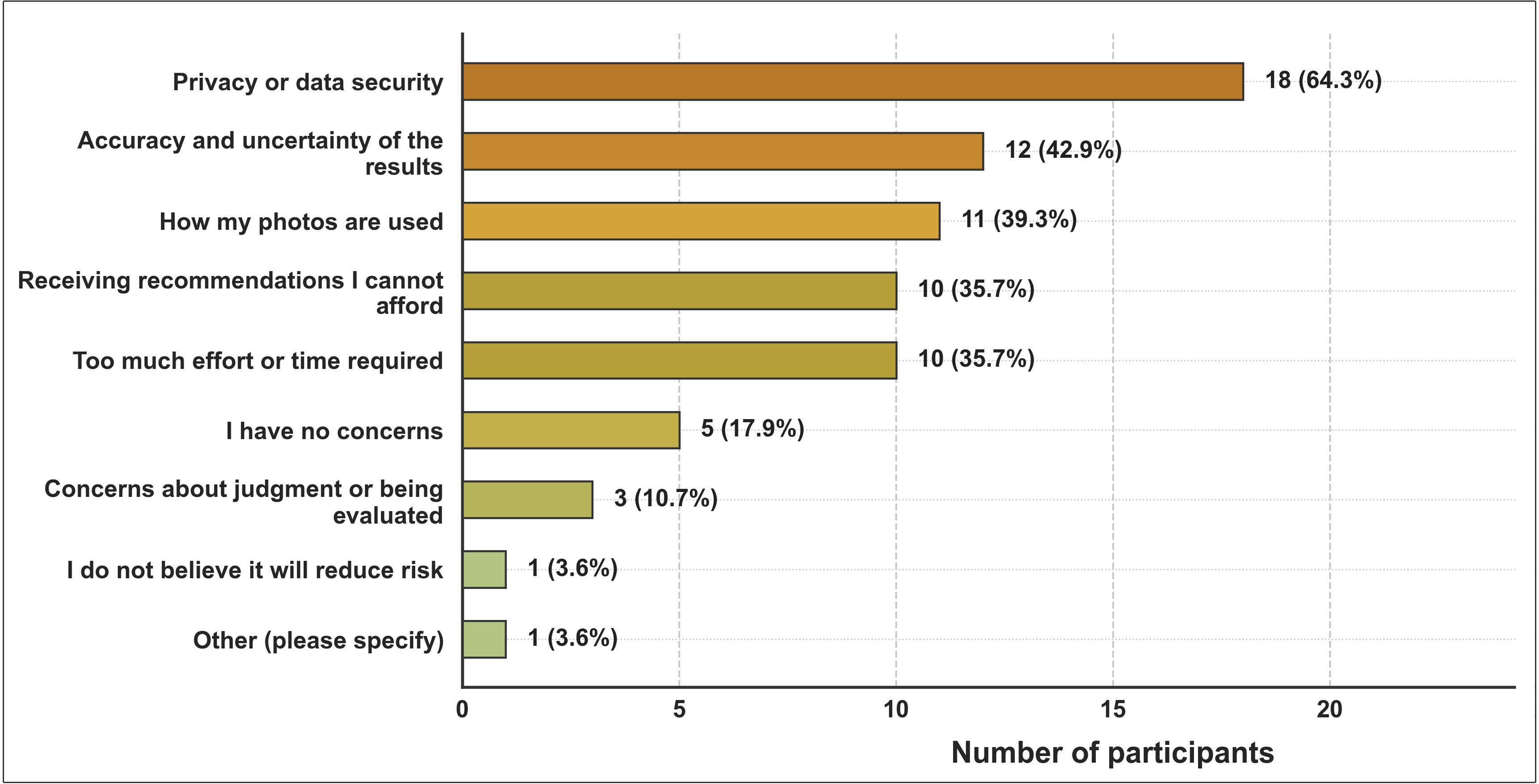}
\caption{Concerns about using the proposed app (Q22; multiple selections allowed; n=28 valid responses). Privacy or data security was the most commonly selected concern (64.3\%), followed by result accuracy or uncertainty (42.9\%), how photographs would be used (39.3\%), unaffordable recommendations (35.7\%), and required time or effort (35.7\%). Percentages exceed 100\% because respondents could select more than one option. Source: Authors’ own work.}\label{fig4}
\end{figure}

Concerns about using the app were widespread. Among the 28 respondents to Q22, 18 (64.3\%) selected privacy or data security, making it the most commonly selected concern. This was followed by result accuracy or uncertainty (12, 42.9\%) and how photographs would be used (11, 39.3\%). The \emph{“I have no concerns”} option was selected by five respondents (17.9\%). Figure \ref{fig4} shows the full distribution. Privacy and accuracy were selected together by 10 respondents, representing 35.7\% of those who answered Q22.

One open-ended survey response captured concern about possible institutional use of the data directly: \emph{``Maybe selling my info to insurance companies.''} Insurance appeared as both a possible benefit and a source of concern. Some participants saw value in using the app \emph{``to keep the house in a condition liked by insurance companies so that I don't get hit by a non-renewal of insurance,''} while others worried that the same information could be used against them. Based on these responses, the proposed tool would need clear and transparent user-controlled restrictions and policies governing how residents’ information could be shared with third-party companies, such as insurers or other institutions.


\subsection{Adoption intent and community sharing}

Participants were generally positive about using the app and sharing its results. Their reported likelihood of using the app if it were available (Q29) averaged 3.77 ($SD = 0.73$), with 66.7\% indicating that they would be likely or very likely to use it. Their likelihood of using the \emph{Save/Share Result} button to keep the result or send it to someone they trusted (Q27) had the same mean of 3.77. Participants' willingness to share their wildfire risk result to support wildfire safety in their community (Q30) averaged 3.70 ($SD = 0.92$), with 53.3\% reporting that they would be willing or very willing to do so.

The association analysis further showed that intention to use the app was connected with saving or sharing the result, acting on app guidance, and sharing information for community wildfire safety. Participants who said they were more likely to use the app also tended to say they would save or share their result ($Q27 \leftrightarrow Q29$, $n = 30$, $\rho = 0.752$, $p_{\mathrm{FDR}} < .001$), act on its recommendations ($Q20 \leftrightarrow Q29$, $n = 30$, $\rho = 0.595$, $p_{\mathrm{FDR}} = .0032$), and share results to support community wildfire safety ($Q29 \leftrightarrow Q30$, $n = 30$, $\rho = 0.601$, $p_{\mathrm{FDR}} = .0029$). In other words, participants who expressed stronger interest in using the tool also tended to see value in using its guidance and sharing the resulting information. 

Participants most often selected family members (73.3\%) and neighbors (56.7\%) as people with whom they would share their results. Smaller proportions selected an insurance provider (23.3\%), a contractor or home-service provider (20.0\%), or the local fire department (16.7\%), while 20.0\% indicated that they would keep the result to themselves. These responses point mainly to household and neighborhood networks, with less interest in sharing results with formal institutions. Although only 20.0\% of participants currently identified neighbors as a source of wildfire-risk information in Q14, 56.7\% said they would share an app-generated result with neighbors. 


\subsection{Practical barriers and conditions for continued use}

\begin{figure}[ht]
\centering
\includegraphics[width=0.90\textwidth]{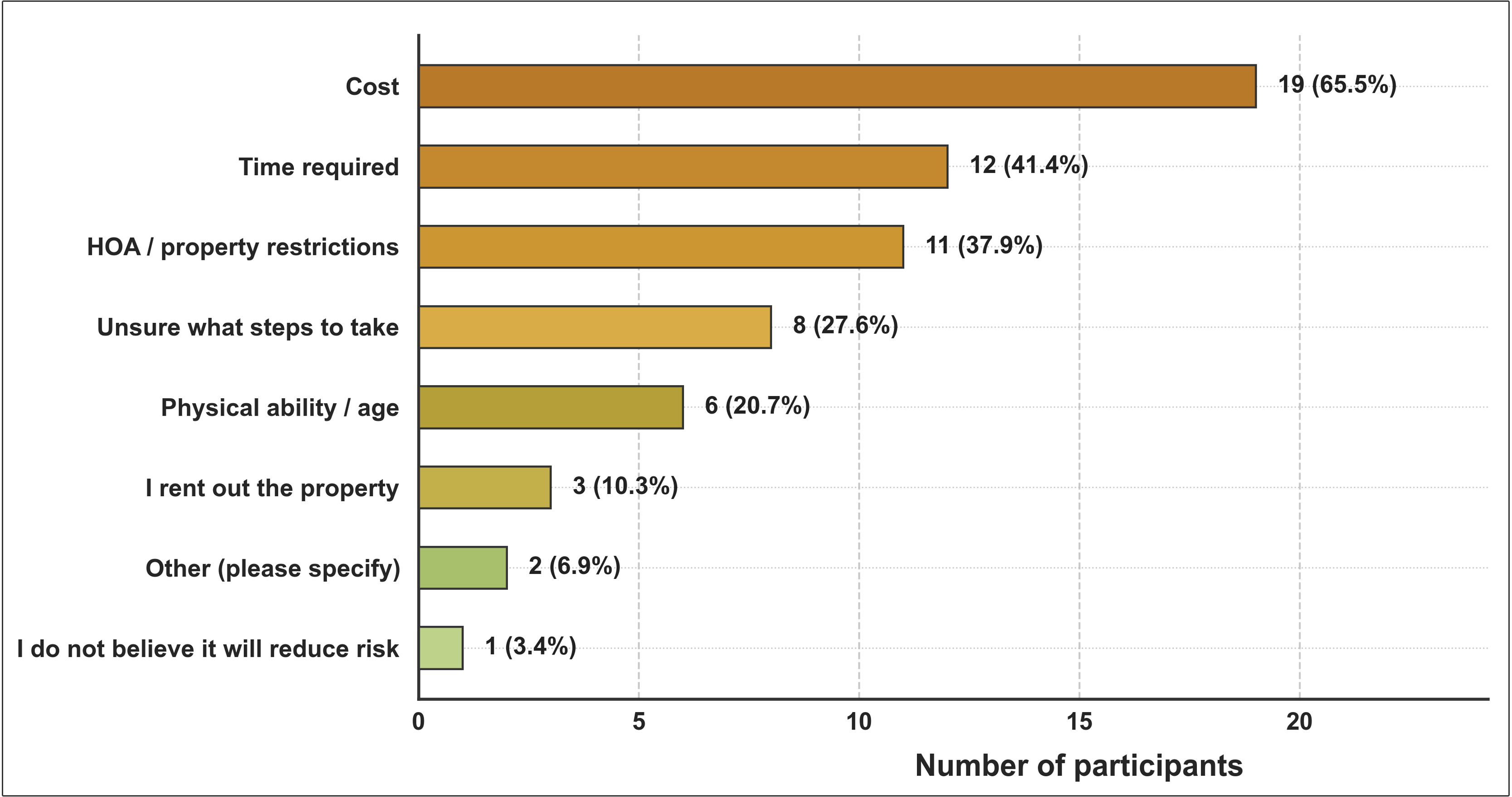}
\caption{Barriers that might prevent participants from acting on app insights (Q21; multiple selections allowed; $n = 29$ valid responses). Cost was the most common barrier (65.5\%), while disbelief that the tool would reduce risk was the least common (3.4\%). Percentages exceed 100\% because respondents could select more than one option. Source: Authors’ own work.}\label{fig5}
\end{figure}

Participants' reported barriers to acting on insights provided by the app were mainly practical (Figure \ref{fig5}). Among the 29 respondents to Q21, cost was selected most often (19, 65.5\%), followed by the time required (12, 41.4\%) and HOA or property restrictions (11, 37.9\%). Eight respondents (27.6\%) were unsure what steps to take, and six (20.7\%) selected physical ability or age. Only one respondent (3.4\%) selected \emph{``I do not believe it will reduce risk.''} Open-ended responses added some details to these barriers. One participant identified \emph{``Disability,''} while another wrote, \emph{``Need to check from several resources to ensure the recommendations are reasonable.''} The latter comment shows that they also wanted to verify the recommendations before acting on them.

The barrier responses also showed that practical constraints frequently overlapped. Among the 29 respondents to Q21, cost and time were selected together by 31.0\%, while cost and HOA or property restrictions were selected together by 24.1\%. These co-selections indicate that cost often appeared alongside the time required to act or limits on what residents were allowed to change. These barriers differed from the privacy and accuracy concerns discussed in Section 5.4 because they concerned whether residents could realistically carry out the recommendations.

According to open-ended Q31 responses about what could make the tool useful over time and encourage continued use, real-time information was mentioned most often, with participants asking for features such as \emph{``If it's like the CalFire website where it shows real-time updates''} and \emph{``Updated risk alerts based on local conditions.''} A second theme focused on safety and community benefit, such as using the tool \emph{``to save us and others from the wildfires.''} A third theme focused on tracking change over time, including \emph{``Tracking my property's risk over time''} and \emph{``Showing how improvements reduce the score.''} A fourth theme was institutional relevance, including whether the tool could be relevant to insurance requirements or supported by trusted partnerships.

Two additional issues also emerged from the qualitative responses. Housing type shaped whether participants felt able to act on recommendations. Comments such as \emph{``If I had a single-family unit''} indicate that residents in apartments (20.0\% of the sample), condominiums (16.7\%), or rental arrangements may not see themselves as having enough authority to make recommended changes. In addition, one participant wanted the app to include evacuation guidance as a complementary feature, noting that it \emph{``could also explain evacuation rules if the unexpected comes before we even get to reduce the risk.''} This points to a possible future link between the tool's prevention focus and active-emergency information.

Across Q25 and Q31 responses, participants were not mainly asking for fewer features. Instead, they wanted clearer connections between the tool's existing parts; from the photo to the score, from the score to the recommendations, from the recommendations to what was feasible, and from the property-level result to the surrounding neighborhood context.


\subsection{Summary of findings}

Overall, the findings show that interest in the proposed tool depended on more than receiving a property-level risk score. Participants saw value in information that could help them understand conditions around their homes, but many were uncertain about mitigation measures already taken, prior inspections, wildfire insurance coverage, or what steps they could take next. The tool would therefore need to support residents with different levels of wildfire knowledge and different levels of control over property-related decisions.

Trust also depended on whether participants could understand how the assessment worked and where its limits were. They wanted to know what the score meant, how visible property conditions affected it, what uncertainty remained, and whether relevant neighborhood factors were considered. Privacy concerns also showed that a clear result would not be enough if residents were still unsure how their photographs and assessment data might be stored, shared, or used by institutions.

Understanding a recommendation also did not necessarily mean that a resident could act on it. Cost, time, HOA or property restrictions, housing type, physical ability, and decision authority affected what participants considered realistic. A recommendation may be clear but still have limited value if it assumes resources or authority that the resident does not have. The AI Value Map in Section \ref{sec:ai_value_map} connects these findings to provisional requirements for the proposed tool.

%% file: sections/06_ai_value_map.tex
\section{The AI Value Map}
\label{sec:ai_value_map}

The \emph{AI Value Map} is the main artifact developed in this paper. It translates the key findings from Section \ref{sec:findings} into traceable design requirements for the proposed AI wildfire risk-assessment tool. The Map organizes participant feedback by connecting each finding to a design requirement and to what may happen if that requirement is not addressed. The Map covers more than screen appearance, navigation, and conventional usability. It includes requirements related to data inputs, model scope, risk-score explanation, uncertainty, privacy, fairness, insurer access, neighborhood context, photo submission, saving and sharing, and whether residents can realistically act on the recommendations. For example, concern about insurer access leads to limits on data sharing; requests to account for neighboring parcels affect what the assessment should consider; and barriers related to cost or housing authority affect how recommendations should be presented. Each row includes four elements: a value dimension identified from the findings, supporting participant evidence, one or more provisional design requirements, and possible consequences if those requirements are not addressed. These requirements have not yet been tested in a functioning system. They are intended to guide the later development and evaluation of the proposed tool.


\subsection{How the AI Value Map was derived}

\begin{figure}[h]
\centering
\includegraphics[width=\linewidth]{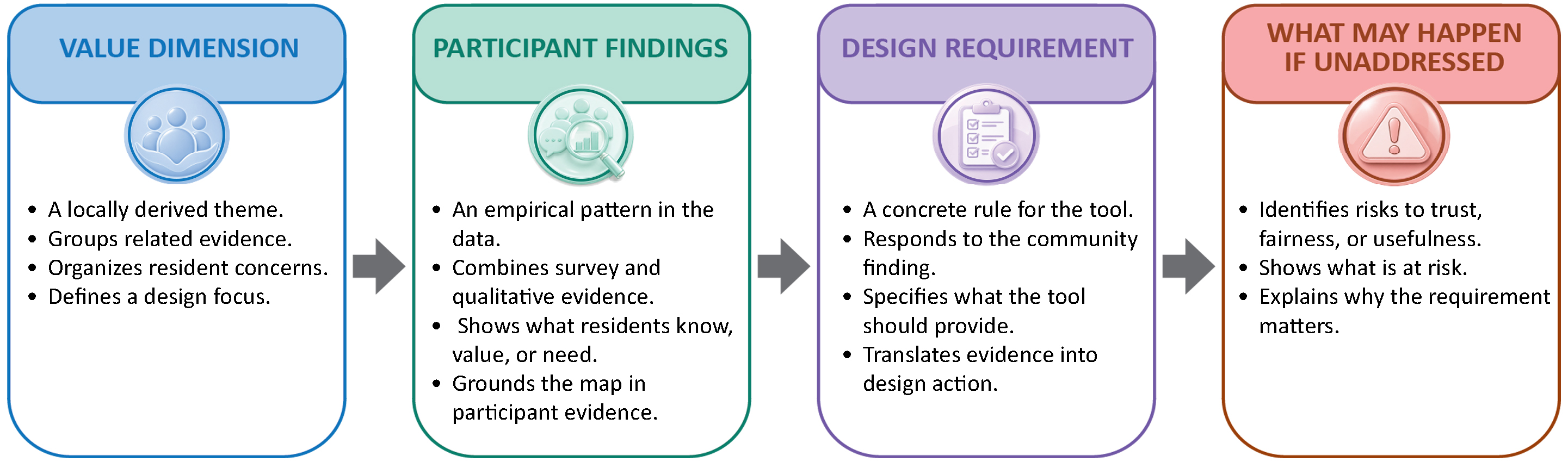}
\caption{Evidence-to-requirement procedure used to construct the AI Value Map. Participant findings are grouped into value dimensions, translated into design requirements, and paired with possible consequences if those requirements are not addressed. The procedure may be adapted to other settings, while the specific dimensions and requirements in this study were derived from the wildfire case. Source: Authors’ own work.}\label{fig6}
\end{figure}

The AI Value Map for this study consists of nine value dimensions. Each dimension represents a recurring participant value, concern, interaction need, or practical constraint that has implications for the proposed tool's design. A dimension is not based on a single survey item or one qualitative theme. Instead, each one brings together related evidence from Section \ref{sec:findings} and connects that evidence to a design response. The map is organized into four columns: \emph{Value Dimension}, \emph{Participant Findings}, \emph{Design Requirements}, and \emph{What May Happen if Unaddressed}. The \emph{Participant Findings} column links each requirement back to the survey and town hall findings. Each value dimension was developed through a five-step procedure. First, related quantitative and qualitative patterns were identified using response distributions, Spearman correlations, open-ended responses to Q25 and Q31, and town hall discussion data. Second, related patterns were grouped into a design-relevant value dimension. Third, the evidence was summarized as a concise participant finding, with participant quotations retained when they clarified concerns not fully captured by the quantitative results. Fourth, each finding was translated into one or more concrete design requirements. Fifth, each requirement was paired with the likely risk to adoption, trust, fairness, or action if it were not addressed. The resulting requirements were developed by the research team from the combined findings and will be refined through later co-design and prototype evaluation. The nine dimensions are ordered broadly around the resident's experience of the tool: how it is framed, how residents provide information, how they interpret results and act on recommendations, how the tool supports continued use, and how fairness, trust, and privacy shape the experience throughout. Figure \ref{fig6} illustrates this evidence-to-requirement procedure, and Table \ref{tab:ai-value-map} applies it across the nine value dimensions.

As an example, the dimension of \emph{privacy boundaries} combines the survey finding that 18 of the 28 respondents to Q22 (64.3\%) selected privacy or data security as a concern with the open-ended comment, \emph{``Maybe selling my information to insurance companies.''} These findings were interpreted as a need for visible data boundaries. The resulting provisional design requirement is a plain-language data policy before photo upload, together with accessible controls to view and delete stored information. Without these safeguards, residents may be less willing to upload property images or return to the tool.


\begingroup
\footnotesize

\makeatletter
\setlength{\LTcapwidth}{\textwidth}
\def\LT@makecaption#1#2#3{%
  \LT@mcol\LT@cols l{%
    \parbox[t]{\LTcapwidth}{%
      \normalfont\small
      \sbox\@tempboxa{#1{\textbf{#2}\quad}#3}%
      \ifdim\wd\@tempboxa>\hsize
        #1{\textbf{#2}\quad}#3%
      \else
        \box\@tempboxa
      \fi
      \endgraf\vskip\baselineskip
    }%
  }%
}
\makeatother

\setlength{\tabcolsep}{4pt}
\setlength{\LTleft}{0pt}
\setlength{\LTright}{0pt}
\renewcommand{\arraystretch}{1.35}

\begin{longtable}{@{}
  >{\centering\arraybackslash}p{0.035\textwidth}
  >{\raggedright\arraybackslash}p{0.170\textwidth}
  >{\raggedright\arraybackslash}p{0.235\textwidth}
  >{\raggedright\arraybackslash}p{0.235\textwidth}
  >{\raggedright\arraybackslash}p{0.255\textwidth}
@{}}

\caption{AI Value Map: nine value dimensions translating participant findings into provisional design requirements}
\label{tab:ai-value-map}\\[3pt]

\toprule
\textbf{\#} &
\textbf{Value Dimension} &
\textbf{Participant Findings} &
\textbf{Design Requirements} &
\textbf{What May Happen if Unaddressed} \\
\midrule
\endfirsthead

\toprule
\textbf{\#} &
\textbf{Value Dimension} &
\textbf{Participant Findings} &
\textbf{Design Requirements} &
\textbf{What May Happen if Unaddressed} \\
\midrule
\endhead

1 &
\textbf{Preparedness Framing}\par
\vspace{0.95\baselineskip}
\textit{how the tool supports preparedness without threat-focused messaging} &
\begin{cellenum}
    \item Wildfire risk concern (2.43/5) and perceived risk (2.41/5) were below midpoint, while preparedness priority was higher (2.93/5).
    \item 33.3\% did not know which mitigation actions had been taken, 36.7\% were unsure about prior inspection, 30.0\% were unsure about wildfire insurance coverage at their properties, and 83.3\% had not experienced wildfire evacuation or property damage.
\end{cellenum} &
\begin{cellenum}
    \item Present the tool as a readiness and prevention help rather than as an alarm or disaster-warning tool.
    \item Explain each recommendation in plain language, including why it matters and what first step the resident can take.
    \item Provide a simple checklist showing what residents know, what they need to verify, and who they may need to contact for help.
\end{cellenum} &
\begin{cellenum}
    \item Threat-focused framing may make the tool feel irrelevant to residents with modest wildfire concern.
    \item Recommendations that assume wildfire knowledge exists may leave residents unsure where to begin.
    \item Without clear guidance, residents may miss unresolved gaps in mitigation, inspection, or insurance coverage.
\end{cellenum}
\\

\midrule

2 &
\textbf{Resident Decision Authority}\par
\vspace{0.95\baselineskip}
\textit{how the tool respects residents' choices and limits on action} &
\begin{cellenum}
    \item Participants saw value in the risk score (3.87) and recommendations (3.80), but were less certain about acting on them (3.50).
    \item The main barriers to action were cost (65.5\%), time (41.4\%), and HOA or property restrictions (37.9\%); 20.7\% also selected physical ability or age.
    \item Housing type affected residents' ability to act: \emph{``If I had a single-family unit.''}
\end{cellenum} &
\begin{cellenum}
    \item Show the cost, time, effort, physical demands, and permissions required for each recommendation.
    \item Organize recommendations from no-cost or low-cost actions to larger investments.
    \item Tailor recommendations to the user's housing situation and decision authority, such as owner, renter, or multi-unit resident.
\end{cellenum} &
\begin{cellenum}
    \item Recommendations beyond a resident's decision authority may seem irrelevant or impossible to carry out.
    \item If feasible and low-cost options are not clearly identified, residents may overlook actions they can realistically take.
    \item Renters and multi-unit residents may disengage if the tool assumes that every user controls a single-family property.
\end{cellenum}
\\

\midrule

3 &
\textbf{Explainability and Result Interpretation}\par
\vspace{0.95\baselineskip}
\textit{what residents need to understand before a risk score becomes meaningful} &
\begin{cellenum}
    \item Color bands were preferred by 53.3\%, while 46.7\% preferred either a numeric score or a descriptive label.
    \item Participants asked what ``72/100 High Risk'' meant and wanted visible highlights connecting features in the uploaded photo to the score.
    \item Participants wanted clearer connections between the identified risk factors, the score, and the recommended actions.
    \item Result accuracy or uncertainty was the second most common app concern (42.9\%).
\end{cellenum} &
\begin{cellenum}
    \item Present the color band, numeric score, and descriptive label together, using the color band as the primary visual cue.
    \item Show a visible breakdown of the factors contributing to the score, with corresponding highlights on the uploaded photo.
    \item Explain in plain language how each factor affects the score and which recommended action addresses it.
    \item Display the uncertainty range and explain its causes and how score changes should be interpreted.
\end{cellenum} &
\begin{cellenum}
    \item A single score format may make the result harder to interpret for residents with different preferences.
    \item Without an explanation of what the score means and what evidence supports it, residents may misread the result or dismiss it as meaningless.
    \item Without a clear connection between identified factors and recommended actions, residents may understand the risk level but still not know what to do next.
    \item If uncertainty is not explained, residents may treat the score as more precise than it is or view changing results as inconsistent.
\end{cellenum}
\\

\midrule

4 &
\textbf{Workflow Guidance and Input Quality}\par
\vspace{0.95\baselineskip}
\textit{how the tool guides photo submission and supports trust in the result} &
\begin{cellenum}
    \item Participants wanted step-by-step photo guidance, clearer instructions at the beginning, reminders to use recent images, and examples of appropriate angles and property coverage.
    \item Comfort with image-based assessment was strongly associated with adoption intention ($\rho = 0.724$).
\end{cellenum} &
\begin{cellenum}
    \item Provide a guided upload process with progress indicators, example photographs, recommended angles, and instructions on property coverage.
    \item Include a pre-submission check for photo recency, clarity, angle, and coverage, and explain that image quality can affect the assessment.
    \item Allow residents to review, replace, or retake photographs before submission.
\end{cellenum} &
\begin{cellenum}
    \item Inadequate upload guidance and quality checks may lead to incomplete or unsuitable photographs and a less dependable assessment.
    \item If residents cannot review or retake photographs, or do not understand why the images are needed, they may abandon the submission or distrust the result.
\end{cellenum}
\\

\midrule

5 &
\textbf{Dynamic Risk Context}\par
\vspace{0.95\baselineskip}
\textit{how the tool stays current and distinguishes photo-based risk from changing external conditions} &
\begin{cellenum}
    \item Participants wanted real-time local updates and alerts.
    \item Town hall participants identified changing or external conditions that a single property photograph may not capture, including wind conditions and human-caused ignition.
    \item They requested evacuation guidance as a complementary feature if conditions changed from preparedness to an active event.
\end{cellenum} &
\begin{cellenum}
    \item Provide current local real-time risk information and update timestamps.
    \item Clearly distinguish photo-assessed property factors from changing external conditions that are not included in the assessment.
    \item Provide links to verified official emergency and evacuation resources, while clearly separating active-event guidance from the tool's property-assessment function.
\end{cellenum} &
\begin{cellenum}
    \item Static or outdated information may make the tool less useful than current services such as CAL FIRE and Watch Duty.
    \item If the assessment scope is unclear, residents may incorrectly assume that the photo-based score accounts for changing external conditions.
    \item Without official emergency links and a clear boundary around the tool's role, residents may rely on it during an active event even though it was not designed for that purpose.
\end{cellenum}
\\

\midrule

6 &
\textbf{Ongoing Use and Return Value}\par
\vspace{0.95\baselineskip}
\textit{what encourages residents to return after the first assessment} &
\begin{cellenum}
    \item Participants wanted to track their property's risk over time and see how mitigation improvements changed the score.
    \item Likelihood of saving or sharing the result was strongly associated with adoption intention ($\rho = 0.752$).
\end{cellenum} &
\begin{cellenum}
    \item Keep a history of past assessments and show changes in the score and contributing factors after mitigation.
    \item Make risk score results easy to save, revisit, and share.
\end{cellenum} &
\begin{cellenum}
    \item A one-time result without assessment history or visible mitigation progress may give residents little reason to return.
    \item If results are difficult to save, revisit, or share, the assessment may remain an isolated output rather than support continued household or neighborhood use.
\end{cellenum}
\\

\midrule

7 &
\textbf{Fairness Across Neighborhoods}\par
\vspace{0.95\baselineskip}
\textit{whether the tool reflects shared risk conditions fairly across neighborhoods} &
\begin{cellenum}
    \item At the concept stage, 76.7\% of participants rated the proposed tool as somewhat or very fair across neighborhoods (mean = 4.13).
    \item Participants linked fair assessment to whether the tool considered neighboring-property maintenance, vegetation on nearby parcels, vacant lots, parcel patterns, and other shared neighborhood conditions.
\end{cellenum} &
\begin{cellenum}
    \item Explain which property and neighborhood conditions contribute to the score, including shared conditions beyond the individual parcel.
    \item Evaluate the tool across different neighborhoods, housing types, and property conditions before making claims about fairness.
\end{cellenum} &
\begin{cellenum}
    \item If shared neighborhood conditions are omitted or unexplained, residents may view the assessment as incomplete or unfair.
    \item Without evaluation across varied settings, systematic differences in relevance or performance across neighborhoods, housing types, or property conditions may become undetected.
\end{cellenum}
\\

\midrule

8 &
\textbf{Trust, Transparency, and Insurance Tension}\par
\vspace{0.95\baselineskip}
\textit{how institutions, sharing, and insurer involvement shape trust} &
\begin{cellenum}
    \item Participants mentioned trusted partnerships and institutional relevance as possible reasons for continued use.
    \item 53.3\% were willing or very willing to share results for community wildfire safety, most often with family (73.3\%) and neighbors (56.7\%).
    \item Insurance appeared as both a possible reason to use the tool and a source of privacy concern.
\end{cellenum} &
\begin{cellenum}
    \item Clearly identify the tool's developer, model limitations, and any genuine institutional or public-agency partnerships.
    \item Give residents control over whether and with whom they save or share their results.
    \item Keep the tool independent from insurers by default, allowing residents to export their own records without automatically connecting data to insurance systems.
\end{cellenum} &
\begin{cellenum}
    \item If the developer, model limitations, or institutional relationships are unclear, residents may question the tool's credibility or whose interests it serves.
    \item Limited sharing control may lead residents to avoid saving or sharing results even when they see value in doing so.
    \item An unclear or automatic connection to insurers may lead residents to fear that their property information could influence insurance decisions or be used against them.
\end{cellenum}
\\

\midrule

9 &
\textbf{Privacy Boundaries}\par
\vspace{0.95\baselineskip}
\textit{how residents control their photos, assessments, and personal information} &
\begin{cellenum}
    \item Among the 28 respondents who answered Q22, privacy and data security were the most common concern (64.3\%).
    \item 39.3\% of Q22 respondents were concerned about how their photographs would be used.
    \item Privacy and accuracy concerns were selected together by 35.7\% of respondents.
\end{cellenum} &
\begin{cellenum}
    \item Before upload, explain how photographs and assessment data will be used, accessed, stored, and deleted.
    \item Give residents clear controls to view and permanently delete stored photographs and assessments.
    \item Do not share photographs, results, or related data with any third party without separate, explicit opt-in consent.
\end{cellenum} &
\begin{cellenum}
    \item Unclear data practices may discourage residents from uploading property photographs or using the tool.
    \item Without accessible viewing and deletion controls, residents may feel that they have lost control of their information.
    \item Unexpected data sharing may reinforce fears that property information could be misused and reduce trust in future assessments.
\end{cellenum}
\\

\midrule
\end{longtable}
\endgroup


\subsection{Key design tensions and considerations}

The following points relate to several AI Value Map dimensions and need attention during the design of the tool.

\textbf{Insurance as both motivation and concern.} Insurance appeared in the findings as both a possible reason to use the tool and a concern. Some participants saw value in documenting property conditions or mitigation progress for their own insurance-related needs. Others worried that the same information could affect their coverage or be used against them. Residents should be able to export and use their own records if they choose. The tool should remain disconnected from insurance systems by default, and photographs, scores, or reports should not be shared with insurers or other third parties without separate and explicit consent. This would allow residents to use the information when it is helpful while keeping control over it.

\textbf{Modest perceived risk alongside preparedness interest.} Participants did not report high levels of wildfire concern or perceived neighborhood risk; however, they placed greater priority on household preparedness and generally saw value in property-specific information and recommendations. The tool should therefore be presented around readiness and prevention, not danger or alarm. It can help residents identify manageable steps, check conditions they are uncertain about, and prepare before an emergency occurs.

\textbf{Context-specific dimensions and a transferable mapping structure.} The nine value dimensions are specific to this exploratory Los Angeles County study and to the conceptual, photo-based wildfire-risk tool examined here. They should not be treated as a fixed checklist for other AI systems. The transferable contribution is the mapping structure: starting with participant evidence, identifying the local design issue and values it reveals, turning that issue or value into a practical design requirement, and showing what may happen if the requirement is not addressed. A team developing a different AI tool could use the same structure, but would be expected to develop different dimensions and requirements based on its own users, setting, and purpose.

%% file: sections/07_discussion.tex
\section{Discussion and Implications}
\label{sec:discussion}

This discussion considers what the AI Value Map and the study findings contribute to community-centered AI, explanation design, and wildfire risk communication. The study does not evaluate a functioning tool; it identifies conditions that a later version of the proposed tool would need to meet.


\subsection{The AI Value Map as a pre-model design method}

The AI Value Map draws on Value Sensitive Design, co-design, and participatory AI. Value Sensitive Design provides methods for examining human values and connecting them to design requirements \citep{FriedmanHendry2019}, while co-design involves intended users in shaping what is developed \citep{SandersStappers2008}. Participatory AI research also warns that community involvement can remain superficial when participants have little effect on important technical or institutional choices \citep{Birhane2022, Sloane2022}. This paper does not claim that considering human values in design is new.

The AI Value Map addresses a practical gap within these approaches. Community findings may be reported clearly, but the path from those findings to specific AI design choices is not always visible. The Map records that path before the model and functional interface are developed. Its contribution is not a new list of values. It is a case-specific procedure for showing how participant evidence becomes a provisional design constraint and what may happen if that constraint is not addressed.

This gives early community involvement a clearer role in the design process. Participants are not only commenting on screen wording or usability. Their concerns can affect what the assessment should consider, how data may be used, and whether the recommendations fit residents' actual circumstances. The Map also allows researchers and designers to examine whether community input has led to a change in the proposed system or has only been documented.

The requirements remain provisional because participants responded to a conceptual tool and no model-generated results were tested. The Map should therefore be treated as a starting point for later co-design, model development, and prototype evaluation. It complements Value Sensitive Design and participatory AI by making the connection between participant evidence and early design choices more explicit.


\subsection{Implications for explanation, actionability, fairness, and governance}

The findings show that explanations for a resident-facing AI tool should do more than identify which features influenced the model result. Participants wanted to understand what the score meant, see how visible property conditions contributed to it, and know what action they could take. Because preferences were divided across color bands, numeric scores, and descriptive labels, the result may need to be presented through several connected forms instead of one fixed format. Methods such as LIME \citep{Ribeiro2016} and SHAP \citep{LundbergLee2017} may support feature-level explanations, but residents also need plain-language interpretation, highlights on the property photograph, uncertainty information, and clear links to recommended actions.

Whether residents can act on those recommendations depends on more than whether they understand them. Cost, time, housing type, HOA or property restrictions, physical ability, and decision authority shaped what participants considered feasible. Later prototype testing should therefore include renters, residents of multi-unit housing, and households with different levels of control over property changes and different financial constraints. The tool may be most useful when it connects general wildfire awareness to conditions at a resident's property and to practical steps that are feasible within that resident's circumstances. This adds a property-level AI design perspective to prior work on wildfire preparedness and risk communication \citep{McCaffrey2015} and on unequal wildfire vulnerability and recovery burdens \citep{Davies2018, MendezFloresHaroZucker2020}.

The findings also show that fairness and trust depend on the scope and governance of the assessment. Participants linked fair assessment to whether the tool considered relevant shared conditions, including wind, neighboring parcels, and vacant lots, not only to whether the same model was applied across neighborhoods. Clear data policies and strict limits on insurer access are similarly important because privacy and insurance concerns may affect whether residents are willing to submit property information. Explanation, feasibility, fairness, and data governance should be treated as connected parts of the resident experience, not as separate design issues.

%% file: sections/08_limitations_future_work.tex
\section{Future Directions}
\label{sec:future_work}

This study provides an initial foundation for understanding residents' values, concerns, and design priorities for a community-centered wildfire risk assessment tool. As an exploratory, pre-model study, the findings are intended to guide the next stages of community engagement, system development, and evaluation rather than provide population-level estimates. The county-wide convenience sample included 30 survey respondents and seven town hall participants and therefore does not support neighborhood-specific comparisons or conclusions about how common the identified perspectives are across Los Angeles County. Although 21 of the 66 tested associations remained significant after FDR adjustment, these results represent exploratory within-sample patterns and should be examined in a larger and more representative sample.

An important next step is to expand engagement within the communities where the proposed tool is intended to be deployed, particularly Altadena, Pasadena, and Pacific Palisades, including residents directly affected by recent wildfires. Such engagement will allow the Value Map to be evaluated and refined against community-specific experiences, priorities, and constraints. In particular, a broader and more diverse sample can provide stronger evidence for refining the \emph{Fairness Across Neighborhoods} dimension and determining whether the identified values and requirements remain consistent across communities with different wildfire exposure, housing conditions, socioeconomic contexts, and capacities for mitigation.

Future studies should also use more consistent data-collection procedures. In the present study, seven town hall participants completed the survey during the live session, whereas the remaining 23 respondents completed it after viewing recorded materials and the presentation. These different formats may have influenced participants' interpretation of the proposed tool and their responses. Subsequent studies can use standardized engagement protocols and larger samples to more systematically examine differences across participant groups and communities.

The next phase of the research will translate the interface and interaction requirements identified through the Value Map into a functional, high-fidelity prototype. Because participants in the present study evaluated wireframes, scenario cards, and a conceptual workflow rather than an operational system, the findings reflect anticipated needs and expectations rather than actual usability or adoption. Prototype evaluation should therefore examine how residents interact with the system under realistic conditions and whether design features such as layered explanations, visible privacy controls, and feasibility-oriented recommendations improve understanding, trust, and willingness to act. Evaluation should include residents with varied housing situations, wildfire experiences, and capacities to implement recommended mitigation measures.

Finally, future work should examine the governance and institutional boundaries surrounding deployment. This includes defining how the tool may interact with insurers, fire agencies, planners, community organizations, and other third parties while preserving resident control over uploaded images and stored assessments. Establishing acceptable data-sharing boundaries, institutional partnerships, and governance mechanisms will be essential before moving from a community-informed prototype toward larger-scale deployment. Together, these next steps will allow the Value Map developed in this study to evolve from an exploratory set of community-derived requirements into a validated framework for the design, evaluation, and governance of a real-world wildfire risk assessment system.

%% file: sections/09_conclusion.tex
\section{Conclusion}
\label{sec:conclusion}

This exploratory study examined how Los Angeles County residents responded to a conceptual AI-driven wildfire risk-assessment tool before a functioning model or operational prototype was developed. Survey and town hall findings showed that participants were cautiously receptive to the concept and recognized its potential benefits. They saw value in property-level risk information and mitigation guidance, but also wanted clear explanations of risk scores, transparent data practices, feasible recommendations, and attention to neighborhood conditions. Preferences for presenting the risk score varied, privacy and data security were common concerns, and practical barriers such as cost, time, housing situation, and decision authority could limit residents' ability to act.

The study's central contribution is the AI Value Map, which gives community input a defined role in early design by connecting participant findings to decisions about the proposed AI assessment and its anticipated outputs, residents' interaction with the tool, and the practical conditions that shape whether its recommendations can be carried out. The findings do not establish that a future tool will be usable, accurate, fair, or widely adopted. They instead identify provisional requirements to guide later design and evaluation. Future work should engage residents more directly in communities previously affected by wildfire, develop and test an operational prototype, and establish clear boundaries for data sharing and institutional partnerships. The AI Value Map offers a resident-informed foundation for these next stages of development and evaluation.